\documentclass[traditabstract]{aa}
\usepackage{graphicx}
\usepackage{txfonts}
\usepackage{natbib}
\usepackage{enumitem}
\usepackage{rotating}
\usepackage{pdflscape}

\bibpunct{(}{)}{;}{a}{}{,}

 \newcommand{\mic}{$\mu$m}

\begin{document}

\title{The emergence of passive galaxies in the early Universe}

\author{
P.~Santini\inst{1}
\and 
 M.~Castellano\inst{1} 
  \and
  E.~Merlin\inst{1}  
  \and
  A.~Fontana\inst{1}   
  \and
  F.~Fortuni\inst{1}
  \and
  D.~Kodra\inst{2}
  \and
  B.~Magnelli\inst{3}
  \and
  N.~Menci\inst{1}
  \and
  A.~Calabr\`o\inst{1}
  \and
  C.~C.~Lovell\inst{4}
  \and
  L.~Pentericci\inst{1}
  \and
  V.~Testa\inst{1}
  \and
  S.~M.~Wilkins\inst{5}
}

  \offprints{P. Santini, \email{paola.santini@inaf.it}}

\institute{INAF - Osservatorio Astronomico di Roma, via di Frascati
  33, 00078 Monte Porzio Catone, Italy
\and Department of Physics and Astronomy and PITT PACC, University of Pittsburgh, Pittsburgh, PA 15260, USA
\and Argelander-Institut f\"{u}r Astronomie, Universit\"{a}t Bonn, Auf dem H\"{u}gel 71, D-53121 Bonn, Germany
\and Centre for Astrophysics Research, School of Physics, Astronomy \&
Mathematics, University of Hertfordshire, Hatfield AL10 9AB
\and Astronomy Centre, Department of Physics and Astronomy, University of Sussex, Brighton, BN1 9QH, UK
}

   \date{Received .... ; accepted ....}
   \titlerunning{The emergence of passive galaxies in the early Universe}

   \abstract{ The emergence of passive galaxies in the early Universe
     results from the delicate interplay among the different physical
     processes responsible for their rapid assembly and for the abrupt
     shut-down of their star formation activity.  Investigating the
     individual properties and demographics of early passive galaxies
     will then improve our understanding of these mechanisms.  In this
     work we present a follow-up analysis of the $z>3$ passive galaxy
     candidates selected by Merlin et al. (2019) in the CANDELS
     fields.  We begin by first confirming the accuracy of their
     passive classification by exploiting their sub-mm emission to
     demonstrate the lack of ongoing star formation. Using archival
     ALMA observations we are able to confirm at least 61\% of the
     observed candidates as passive. While the remainder lack
     sufficiently deep data for confirmation, we are able to validate
     the entire sample in a statistical sense.  We then estimate the
     Stellar Mass Function (SMF) of all 101 passive candidates in
     three redshift bins from $z=5$ to $z=3$. We adopt a stepwise
     approach that has the advantage of taking into account
     photometric errors, mass and selection completeness issues, and
     the Eddington bias without any a-posteriori correction. We
     observe a pronounced evolution in the SMF around $z\sim4$,
     indicating that we are witnessing the emergence of the passive
     population at this epoch. Massive ($M>10^{11}M_\odot$) passive
     galaxies, only accounting for a small ($<10\%$) fraction of
     galaxies at $z>4$, become dominant at later epochs. Thanks to a
     combination of photometric quality, sample selection and
     methodology, we overall find a higher density of passive galaxies
     than previous works. The comparison with theoretical predictions,
     despite a qualitative agreement (at least for some of the models
     considered), denotes a still incomplete understanding of the
     physical processes responsible for the formation of these
     galaxies.  Finally, we extrapolate our results to predict the
     number of early passive galaxies expected in surveys carried out
     with future facilities.  }

\keywords{Galaxies: evolution - Galaxies: high-redshift - Galaxies:
  luminosity function, mass function - Methods: data analysis}

\maketitle

\section{Introduction}\label{sec:intro}

It has become evident, in the last 10-15 years, that massive passive
galaxies were already in place at earlier and earlier epochs
\citep[e.g.][]{cimatti04,saracco05,labbe05,kriek06,grazian07,fontana09,straatman14,nayyeri14,schreiber18c,merlin18,merlin19,carnall20,shahidi20}. The
very existence of massive galaxies with suppressed star formation at
$z>3$ poses serious challenge to theoretical predictions, that
struggle to reproduce intense enough star formation rates (SFR) at
even higher redshift to assemble such large stellar masses
($>10^{10}-10^{11} M_\odot$), and prevent further gas collapse at
least until the epoch of observation
\citep{fontana09,vogelsberger14,feldmann16,merlin19,shahidi20}. Despite
the difficulties associated with their search, substantially due to
their faintness in the UV rest frame commonly used to select
high-redshift galaxies, as well as with their spectroscopic
confirmation, studying and deeply understanding these systems will
help us shed light on their rapid formation process and similarly
rapid quenching mechanism.

This paper is the fourth in a series. In our first work
\citep{merlin18} we presented an accurate and conservative technique
to single out passive galaxies at high redshift by means of SED
fitting with a probabilistic approach. We selected 30 $z>3$ candidates
in the GOODS-S field.  Passive galaxy candidates, while being
relatively easy to select from photometric surveys once the technique
is established, need to be confirmed by other means. This is usually
achieved through spectroscopic observations. However, spectroscopy
becomes particularly difficult and time consuming at $z>3$, where only
few candidates have been confirmed so far
\citep{glazebrook17,schreiber18c,tanaka19,valentino20,forrest20,forrest20b,saracco20,deugenio21}.
In our second work \citep[][S19 hereafter]{santini19}, we used a
complementary approach, and looked for evidence of lack of star
formation as seen in the sub-mm regime to confirm the passive
classification of the high-$z$ candidates selected in GOODS-S.  At
that time, we could confirm 35\% of the targets on an individual basis
adopting conservative assumptions, and validated the sample as a whole
in a statistical sense.  In our third work \citep[][M19
hereafter]{merlin19}, we extended the search for passive galaxies to
the entire CANDELS sample, and selected 102 $z>3$ candidates over the
five fields.  In the present work, we first confirm the passive nature
of these candidates adopting the method presented in S19 and taking
advantage of the richer ALMA archive (that includes observations that
were still proprietary at the time of our previous work). We then
analyse the emergence and mass growth of this peculiar class of
galaxies by means of two powerful statistical tools: the Stellar Mass
Function (SMF) and the Stellar Mass Density (SMD).  Very few studies
so far have pushed the analysis of the SMF of passive galaxies at
$z>3$ \citep{muzzin13,davidzon17,ichikawa17,girelli19}, because of the
difficulty in assembling statistically meaningful samples of
candidates at such high redshift.

The paper is organized as follows. We present the data set and recall
how passive galaxies are selected in Sect.~\ref{sec:data}; in
Sect.~\ref{sec:alma} we confirm the passive nature of the sample by
means of ALMA public observations; Sect.~\ref{sec:mfdet} illustrates
the method adopted to compute the SMF; our results are presented in
Sect.~\ref{sec:mf} and compared with previous observations and
theoretical predictions in Sect.~\ref{sec:comparison}; based on our
SMF, in Sect.~\ref{sec:predictions} we predict the number of high-$z$
passive galaxies expected from future facilities; finally, we
summarize the main findings of this work in Sect.~\ref{sec:summary}.
In the following, we adopt the $\Lambda$ Cold Dark Matter (CDM)
concordance cosmological model ($H_0 = 70$ km s$^{-1}$ Mpc$^{-1}$,
$\Omega_M = 0.3$, and $\Omega_\Lambda = 0.7$) and a \cite{salpeter55}
Initial Mass Function (IMF). All magnitudes are in the AB system.

\section{The data set and sample selection}\label{sec:data}

We briefly summarize here the major characteristics of the CANDELS
data set and the strategy adopted to select passive galaxies, and
refer the reader to the relevant publications for further details.

The Cosmic Assembly Near-infrared Deep Extragalactic Legacy Survey
(CANDELS; \citealt{koekemoer11}, \citealt{grogin11}; PIs: S. Faber and
H. Ferguson) is the largest (902 orbits) HST program ever approved,
and it has observed the distant Universe with the WFC3 and ACS
cameras. It has targeted 5 fields, covering a total area of $\sim$
1000 sq. arcmin. The multiwavelength photometric catalogues have been
publicly released and are fully described in the relevant accompanying
papers (\citealt{guo13} for GOODS-S, \citealt{barro19} for GOODS-N,
\citealt{galametz13} for UDS, \citealt{stefanon17} for EGS and
\citealt{nayyeri17} for COSMOS).  For GOODS-S we used an improved
catalogue including three more bands (WFC3 F140W, VIMOS $B$ and
Hawki-$K_s$ from the HUGS survey, \citealt{fontana14}) and improved
photometry on the Spitzer bands thanks to new mosaics (IRAC CH1 and
CH2, by R. McLure) and new software (all four channels were
re-processed using T-PHOT, \citealt{merlin15}). These improvements are
discussed in \cite{merlin21}.  For the remaining fields we used the
official catalogues.

We adopted the photometric redshifts from the latest CANDELS estimates
\citep{kodraphd19}, to be presented in Kodra et al. (in prep.). These
improve upon the original released ones presented in \cite{dahlen13},
and are based on a combination of four independent photo-$z$
probability distribution functions (unless spectroscopic redshifts are
available), using the minimum Frechet distance combination method.

\subsection{The selection of high-$z$ passive candidates} \label{sec:passivesel}

The present work is based on the $z>3$ passive galaxy sample selected
by M19 in the five CANDELS fields.  Candidates have been identified by
means of an {\it ad-hoc} developed SED fitting technique, fully
described in \cite{merlin18}. Briefly, we used \cite{bc03} stellar
population models and assumed ``top-hat'' star formation histories
(SFH), characterized by a period of constant star formation followed
by an abrupt truncation of the star formation, that is set to zero
thereafter. Despite being over-simplified, these analytic shapes
manage to mimic the rapid timescales available for quenching at high
redshift, comparable to the age of the Universe (at variance with
standard exponential models, that require at least $\lesssim$1 Gyr to
reach quiescence). Moreover, the adoption of ``top-hat'' SFHs improves
over a standard $UVJ$ selection since the fit is able to identify
recently quenched sources, still showing blue $U-V$ colours.

In addition of being fit in their passive phase (i.e. with zero SFR),
candidates also need to pass conservative selection criteria based on
the probability $P$($\chi^2$) of the $\chi^2$ resulting from the
fitted solution. In the end, in order to be classified as passive,
galaxies need to fulfill the following criteria:
\begin{itemize}
\item $H<27$;
\item 1$\sigma$ detection in $K$, IR1 and IR2 bands;
\item $z_{phot}>3$;
\item SFR$_{best}$=0
\item $P$($\chi^2_{Q}$)$>$30\% {\it and} $P$($\chi^2_{SF}$)$<$5\%.
\end{itemize}
The last requirement ensures that the best-fit passive solution has a
high probability $P$($\chi^2_{Q}$) and at the same time that no
plausible (i.e. with $P$($\chi^2_{SF}$)$>$5\%) star-forming solutions
exist.

\begin{figure}
  \centering
 \includegraphics[width=\hsize]{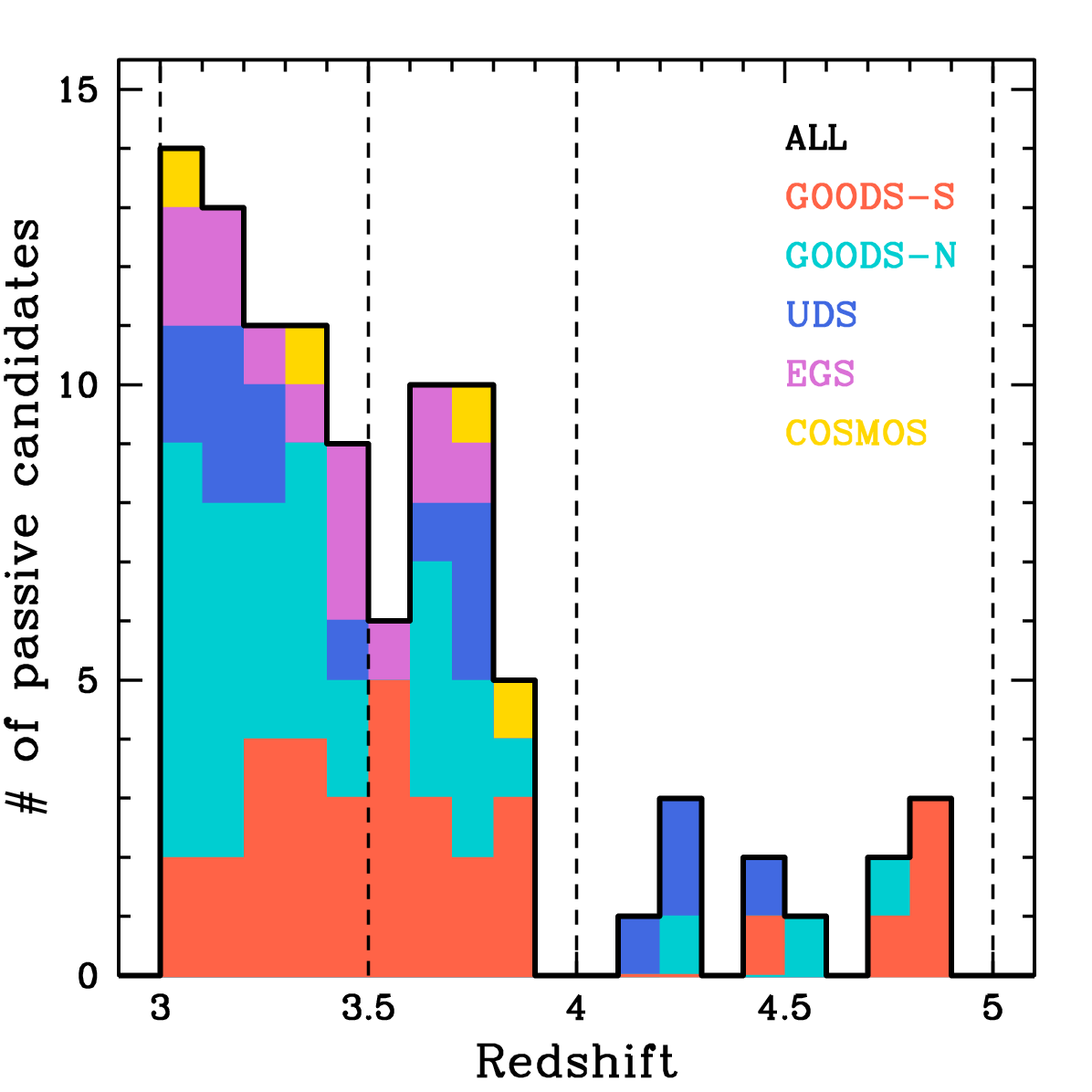}
 \caption[]{Redshift distribution of the $3\leq z < 5$ passive
   candidates used in the present analysis. Different colours show the
   counts in the various fields (see legend). The black open histogram
   shows the total sample.}
              \label{fig:zdist}
\end{figure}

We ended up with a sample of 102 galaxies (the so-called ``reference''
sample in M19, i.e. that obtained without inclusion of emission lines
in the fit). Candidates are distributed quite inhomogeneously across
the fields: 33 in GOODS-S, 36 in GOODS-N, 16 in UDS, 13 in EGS, 4 in
COSMOS. The variance across the fields cannot be entirely explained by
cosmic variance, but probably reflects the different photometric
properties of the five catalogues.  In particular, they differ in the
depth of the $K$ and IRAC bands (see M19). All candidates have
redshift lower than 5 except one source in GOODS-N
($z_{phot}\sim 6.7$, discussed in M19). In the following, we restrict
to the $3<z<5$ sample, made of 101 galaxies.  The final list of
passive candidates is presented in Tables~\ref{tab:gs} to
\ref{tab:egs}.

\begin{figure*}
    \centering
 \includegraphics[width=0.85\columnwidth,angle=270]{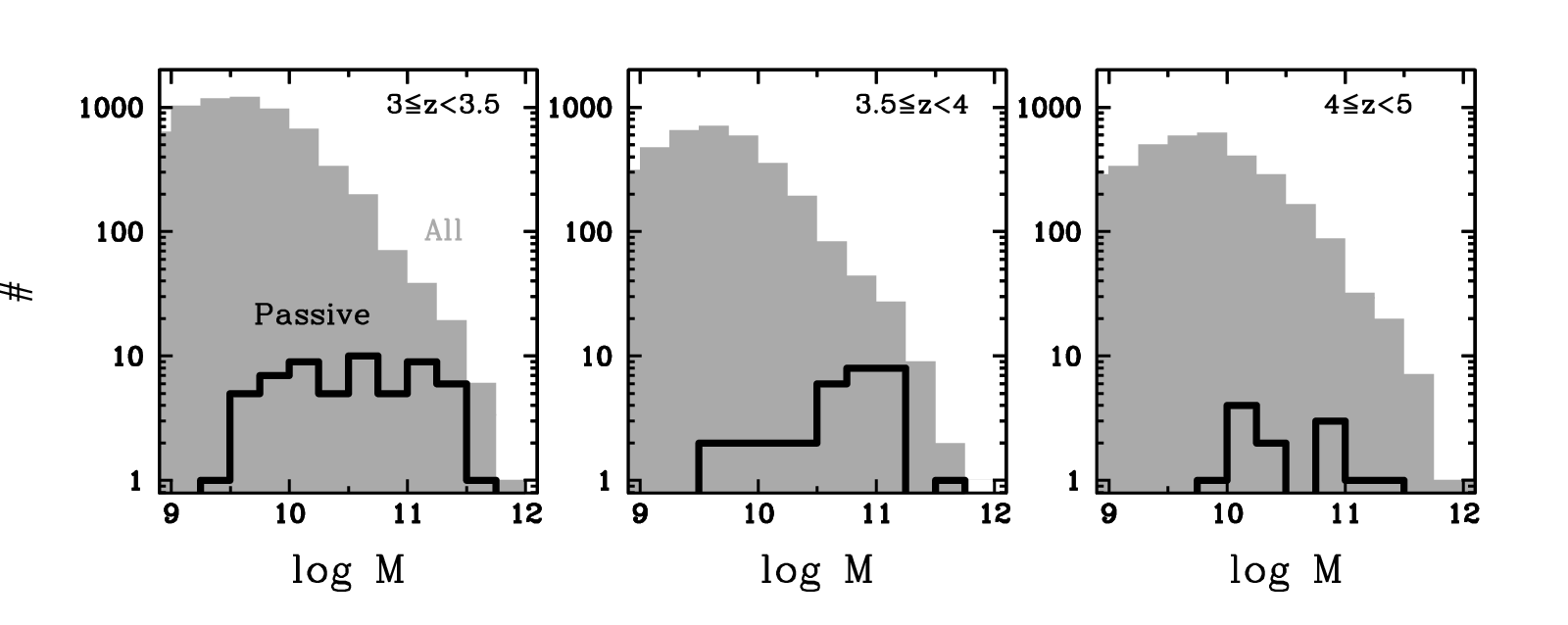}
 \caption[]{Mass distribution in the passive galaxies (black open
   histograms) compared to the overall galaxy population (gray shaded
   ones) in different redshift bins.  }
              \label{fig:mdist}
\end{figure*}

Among our passive galaxies, 4 are flagged as X-ray detected AGNs in
GOODS-S (one of which is a confirmed AGN, see M19), 3 in GOODS-N and 2
in EGS. Two of the AGN candidates in GOODS-S are also detected by {\it
  Herschel} \citep[see][]{santini19}, likely due to the emission of
the host dust heated by the central nucleus. We believe the AGN
candidates are therefore obscured, or at least not bright Type 1 AGNs.
AGN emission does not significantly affect the estimate of the stellar
mass for obscured AGNs, as demonstrated by \cite{santini12b}. For this
reason, we did not exclude AGNs from the sample.

Fig.~\ref{fig:zdist} shows the redshift distribution of our passive
candidates, colour-coded according to the field they belong to. The
majority of them are below redshift 4, but the distribution shows a
non negligible tail to higher redshift (12 galaxies at $4<z<5$).

\subsection{Stellar masses} \label{sec:stellarmasses}

Stellar masses have been estimated by SED fitting the observed
near-UV-to-near-IR photometry, up to 5.5$\mu$m rest-frame, to
libraries built on \cite{bc03} stellar population synthesis models.

For passive galaxies they have been obtained from the fit with
``top-hat'' SFHs without including emission lines, as discusses in
Sect.~\ref{sec:passivesel}. The grid in stellar ages, burst duration,
metallicities and dust extinction is detailed in M19 (see their
Appendix A).

To test the reliability of the inferred stellar masses we explored the
possibility that our candidates host an on-going dusty burst,
misinterpreted as an old stellar population in our fits. We therefore
built a stellar library by combining an old, unobscured or moderately
obscured burst finished by at least 100 Myr with a star formation
episode on-going since 50 Myr. We fit our GOODS-S candidates with this
2-component library and found very similar stellar masses, with a
median $M_{tophat}/M_{2comp}$ of 0.96 and semi interquartile range of
0.12. We also found that in 29 out of 33 candidates the fit prefers
solutions with purely old stellar populations, and that in the
remaining 4 sources the young stellar population contributes by no
more than 20\% in mass. This very same result is found by fitting mock
galaxies that are purely passive, indicating that in a small fraction
of objects (13\%) the fit is limited by parameter
degeneracy. 

For the whole galaxy population we adopted exponentially declining
models ($\tau$-models) to parameterize the SFH. The library has been
built as done in previous works
\citep[e.g.][]{fontana04,santini09,grazian15}: metallicities range
from $Z = 0.02 Z_\odot$ to $Z = 2.5 Z_\odot$; $0 < E(B-V) < 1.1$ with
a \cite{calzetti00} or Small Magellanic Cloud \citep{prevot84}
extinction curve, left as a free parameter; the timescale for the
exponentially declining SFH ranges from 0.1 to 15 Gyr; the age
(defined as the time elapsed since the onset of star formation) varies
within a fine grid and, at a given redshift, is set lower than the age
of the Universe at that redshift.

The mass distributions of passive candidates and of the whole
population in the redshift intervals used in the present analysis are
shown in Fig.~\ref{fig:mdist}.

While being very effective in selecting quiescent galaxies at high
redshift, the adoption of ``top-hat'' models does not have a strong
impact on stellar masses. This confirms the results of
\cite{santini15}, who found that the stellar masses are stable against
the choice of the SFH parameterization and differences in the
metallicity/extinction/age parameter grid sampling. However, as we
discussed in our previous work, while the stellar mass is on average a
relatively stable parameter, it can vary for some peculiar sub-classes
of objects for which a different SFH parameterization is necessary,
such as passive galaxies at high redshift.
 
\begin{figure}
    \centering
   \includegraphics[width=0.66\columnwidth,angle=270]{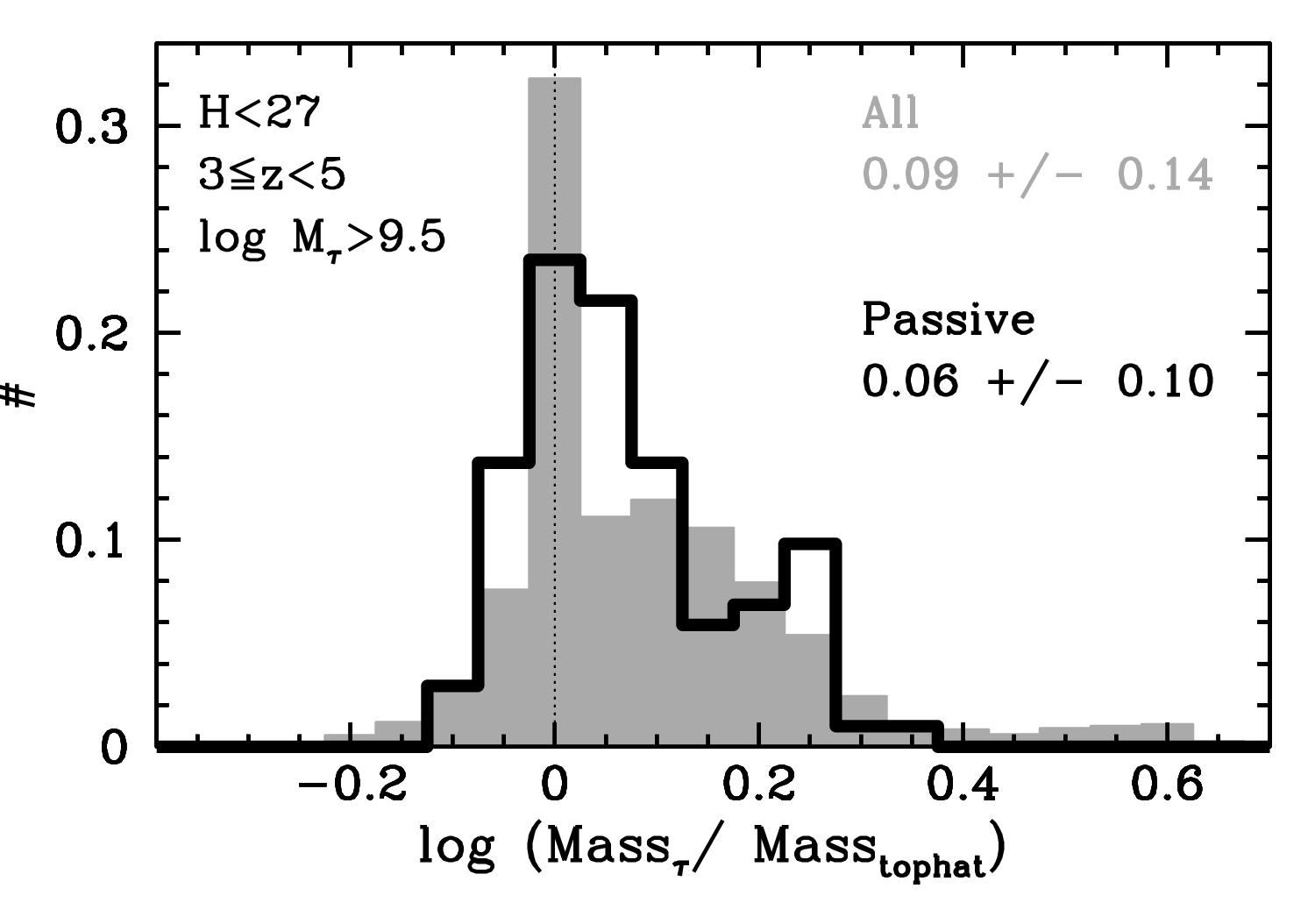}
   \caption[]{Distribution of the ratio between the stellar mass
     computed with $\tau$-models and with ``top-hat'' SFHs. The gray
     shaded histogram refers to the total population of $H<27$
     galaxies at $3\leq z<5$ with $M_\tau>10^{9.5}M_\odot$ in the five
     CANDELS fields, while the black open one shows the passive
     candidates. The numbers in the upper right corner show the mean
     and standard deviation of the two distributions. }
              \label{fig:tau_th}
\end{figure}

Fig.~\ref{fig:tau_th} shows a comparison between stellar masses
obtained with $\tau$-models and with ``top-hat'' models, for the whole
population in the redshift range of interest (gray shaded histogram)
and for the passive candidates (black open histogram). Despite being
slightly skewed towards larger $M_\tau /M_{tophat}$ ratios
($<\log{M_\tau /M_{tophat}}>=0.06$ and 0.09 for the passive and global
population, respectively), the distributions are peaked at unity, and
the mean value is well within the standard deviation of the
distribution (0.10 and 0.16, respectively). The latter is also smaller
than the average error bar associated with the estimate of the stellar
mass ($\sim$0.4 dex for the entire $M>10^9M_\odot$, $\sim$0.2 dex at
$M>10^{10}M_\odot$). We conclude that the two stellar mass estimates
are in good agreement for the large majority of sources. A systematics
however appears for the passive candidates, whose distribution of
$\log{M_\tau /M_{tophat}}$ shows a secondary peak at $\sim$0.2
dex. This feature is entirely due to candidates with stellar mass
below $10^{10.5}M_\odot$, that show a broad (or slightly bimodal)
distribution between 0 and 0.2 dex. The objects showing the largest
mismatch ($\gtrsim$0.1 dex) have been fitted with implausibly low
values for the stellar metallicity (0.02 times Solar) in the
$\tau$-model library, which was built for the global galaxy population
and not optimized for passive sources. For this reason, in the
following, when computing the SMF for the entire galaxy population, we
replaced the stellar mass values of the passive candidates with the
results obtained from the ``top-hat'' library, that we consider more
appropriate for this class of objects.

\section{Confirmation of the passive nature of the candidates from their sub-mm emission}\label{sec:alma}

One powerful possibility to validate the passive classification of
photometrically selected candidates is to measure their FIR and/or
sub-mm emission, used to exclude on-going star formation. This method
allows us to avoid possible degeneracies between dusty star-forming
and old passive solutions in the optical fit.  In M19 we cross-matched
our sample with Herschel catalogues. For the GOODS fields we took
advantage from the new, deep ASTRODEEP catalogues by Wang et al. (in
preparation), which combine data from PEP \citep{lutz11} and
GOODS-Herschel \citep{elbaz11} surveys for the PACS bands and from the
HerMES survey \citep{oliver12} for the SPIRE bands. For COSMOS and EGS
we used PEP-PACS catalogues and HerMES-SPIRE ones \citep{roseboom10,
  roseboom12}. For the UDS field we used HerMES catalogues for both
cameras.  As discussed in our previous work, with the exception of two
sources in GOODS-S (GOODSS-10578 and GOODSS-3973), which are likely
obscured AGNs (see discussion in \citealt{merlin18}, S19, M19 and in
Sect.~\ref{sec:validation}), no clear evidence for FIR emission was
found. This test is however limited by poor sensitivity at these
redshifts and high level of confusion. For this reason, we attempted
to extract more information out of Herschel maps by stacking.  We
stacked all sources at 100 and 160\mic, offering the best compromise
between depth and angular resolution among all Herschel bands. We did
not measure any flux above the noise level in any of the five fields
individually nor in the combination of them. We note, however, that
the depth of stacked Herschel maps, resulting into a typical limiting
SFR of $\sim$100 (600) M$_\odot$/yr from 100\mic~(160\mic), is not
sufficient to reject mild levels of SFR in these galaxies. For this
reason we opted for ALMA data to confirm the passive solutions of our
candidates.

We took advantage of the rich ALMA archive, and extended the very same
analysis described in S19 to the entire CANDELS sample of high-$z$
passive candidates. In the following, we summarize the main steps and
present the new results. We refer the reader to S19 for further
details.

\subsection{ALMA archival observations} 

Out of the five CANDELS fields, only three are accessible by ALMA
(GOODS-S, UDS and COSMOS). This reduces the number of candidates over
which the validation method can be attempted to 53. A search in the
ALMA archive returned observations in Band 6 and 7 for 41 of them.
Nine candidates in UDS and three candidates in COSMOS were observed in
Band 7, with one of the latter also observed in Band 6. 21 and 19
candidates in GOODS-S have been observed in Band 6 and 7,
respectively, with 11 observed in both bands.  Band 3 and 4
observations, available for a fraction of the sources, give less
stringent constraints and have not been used.

We produced images with $\geq$0.6 arcsec spatial resolution using
``natural'' weightings and $uv$-tapers when needed. In this way, we
could assume that sources are unresolved (see justification and
references in S19), and maximize the sensitivity to detection. We note
that given the native angular resolution of the data, the maximum
recoverable scale is $>$1.5 arcsec (0.75 arcsec in a couple of
cases). Assuming an homogenous surface brightness, this results into a
physical scale of $>$10 kpc ($\sim$5 kpc) at the redshifts of our
sources.  The typical size of $z>2$ sources measured by
\cite{fujimoto17} on ALMA Band 6 and 7 images ($R_e$$<$3
kpc, with an avearage value of $\sim$1
kpc) makes us confident that the emission of our targets is not
filtered-out by the reduction process.

The sub-mm flux was read directly on the map (in units of Jy
beam$^{-1}$)
as the value of the pixel at the position of the source. The
associated uncertainty was taken as the standard deviation of all
pixels with similar coverage, after excluding the position of the
candidate and potential other sources in the image (see S19).  In
order to avoid astrometry issues, if the source is detected at more
than 3$\sigma$,
we replaced the value of the flux at the centre with the maximum value
within the beam; if, on the other hand, the source is undetected, we
used the mean flux within the beam.  We finally stacked sources
observed more than once in the same Band by averaging the fluxes
measured from different programs weighting them with the associated
errors (the final sensitivity per beam was inferred as the standard
error on the weighted mean).

The achieved median sensitivity is 0.17 mJy in Band 6 and 0.25 mJy in
Band 7 (with a semi interquartile range of scatter among the various
sources of $\sim$0.07 in both cases).  Only one of the 41 sources is
detected at $>$3$\sigma$ level (GOODSS-3973,
$S_{870\mu m}=0.74\pm0.18$ mJy). Six sources are barely above the
noise level (at 1-2 $\sigma$ level). We report in Tables~\ref{tab:gs}
to \ref{tab:cosmos} the ALMA flux in the most stringent band.  No
significant detection is found even after stacking all observations in
the same Band (including the only detected source), either over the
entire redshift range or in bins of redshift ($3<z<4$ and $4<z<5$).

\subsection{Method}  \label{sec:sfr}

Sub-mm fluxes were converted into estimates of (or in most cases upper
limits on) the SFR by fitting them to a number of IR models. We
considered the average SMG SEDs of \cite{michalowski10} and
\cite{pope08}, the two average SEDs fitted by \cite{elbaz11} for Main
Sequence and starburst galaxies, and the full libraries of
\cite{ce01}, \cite{dh02} and \cite{schreiber18}. The latter library
has the dust temperature $T_d$ as free parameter, whose choice highly
affects the inferred SFR. We considered two templates, one with
$T_d=25$K, as conservatively expected for quiescent galaxies
\citep{gobat18,magdis21} and one with light-weighted $T_d=40-50$K, as
predicted by the redshift evolution of the dust temperature
parameterized by \cite{schreiber18}. We assumed
$R_{SB} = $SFR$/$SFR$_{MS} = 1$. As done in S19, in the following we
adopted the template of \cite{michalowski10} as reference model. We
converted the inferred total infrared luminosity between 3 and 1100
$\mu$m adopting the calibration of \cite{kennicutt12}, adjusted to a
Salpeter IMF using their conversion factor. The SFR was calculated at
the redshift of the candidates and at any given redshift between 0 and
6. For the majority (10 out of 12) of the candidates observed both in
Band 6 and 7, we used the former, as the data turned out to be deeper
in terms of the resulting SFR. The median limiting SFR is $\sim$43
M$_\odot$/yr ($\sim$55 M$_\odot$/yr at $z>4$), with a semi
interquartile range of $\sim$15 M$_\odot$/yr.  The inferred SFRs are
listed in Tables~\ref{tab:gs} to \ref{tab:cosmos}.  We note that any
possible contribution from evolved stellar populations to the FIR
luminosity would decrease the inferred SFR, and therefore would make
our results even stronger.

As discussed in S19, the spread in the inferred SFR among different
templates is not negligible (it may span even a decade when fitting to
the longest wavelength Band 6 data). As a consequence, the exact value
of the SFR is subject to this systematics. However, although we
discuss the results based on the reference model of
\cite{michalowski10}, we note that our conclusions are solid against
the choice of the IR templates.
 
We used two approaches to validate the passive solutions of the
optical fit.

\subsubsection{Validation of individual candidates} \label{sec:validation}

To make sure that our candidates were not erroneously best fitted by
passive templates, we checked whether the alternative star-forming
solutions are compatible or not with ALMA results. To this aim, we
compared the SFR inferred from ALMA data, or in most cases the
3$\sigma$ limits on the SFR, as a function of redshift to any
plausible (i.e. with associated $\chi^2$ probability larger than 5\%)
star-forming solutions of the optical fit obtained at any
redshift. The fit was indeed re-run with redshift set as a free
parameter, instead of fixed to the best-fit one (where, by definition,
there are no star-forming solutions with probability larger than 5\%).

For 61\% of the observed candidates (including the only detected
source), ALMA constraints predict a SFR that, at any redshift, is
below the optical solution (or in most cases no star-forming solutions
exist at all). In other words, the very red colours of the candidates
cannot be justified by a large amount of dust, but need to be
explained by old stellar populations. This implies that the optical
star-forming solutions are implausible, therefore supporting the
passive best-fit. These sources are robustly and individually
confirmed as passive at a 3$\sigma$ confidence level, and the result
is unaffected by the choice of the IR template used to calculate the
SFR (we note that even more sources would be confirmed adopting more
conservative templates).  The last column of Tables~\ref{tab:gs} to
\ref{tab:cosmos} indicates the candidates that have been individually
confirmed.  For the rest of the sample, ALMA upper limits are above
the optical star-forming solutions, making the comparison
inconclusive. Deeper sub-mm data would be needed to confirm (or
reject) the passive nature of these sources.  This comparison is
graphically represented in Fig. 3 of S19 for the GOODS-S candidates of
\cite{merlin18}.  The improvement in the fraction of validated
candidates compared to S19 results is explained by two effects. In
part, the adoption of updated redshifts for the search of passive
galaxies leads to a more reliable selection. Most importantly, our
previous work relies on the exceedingly conservative choice of
adopting, for this test, the nebular line fit for all candidates, even
for those selected without their inclusion.

Among the candidates that have been individually confirmed as passive
are the two GOODS-S candidates GOODSS-10578 and GOODSS-3973 with
associated Herschel (and MIPS) emission, the latter also detected by
ALMA. Though it may at first be surprising, as fully discussed in
\cite{merlin18}, S19 and M19, these two sources are likely obscured
AGNs, both of them being identified as X-ray emitters in the {\it
  Chandra} catalog of \cite{cappelluti16}, and the former also
confirmed as AGN by the MUSE deep data \citep{inami17}. Their IR
emission is indeed likely associated with hot dust heated by the
central AGN rather than cold dust heated by UV stars and tracing
on-going star formation. As a matter of fact, according to the IR
templates the we used to estimate the SFR (see Sect.~\ref{sec:sfr} and
S19), the mid-IR emission of these two sources is at least a factor of
10 (if not 100, depending on the chosen model) lower than expected
based on ALMA flux.  Similarly, the recent analysis of
\cite{deugenio21} found four secure 24\mic~detections out of their 10
$z\sim 3$ spectroscopically confirmed passive galaxies, with no FIR
counterparts, three of which consistent with being AGN-powered. Based
on a mean stack radio detection they derived similar limiting SFR as
individually measured by us from ALMA data. Furthermore,
\cite{spitler14} found six 24\mic~detections among their 26 massive
quiescent (based on a UVJ selection) $3<z<4$ candidates, three of
which showing signs of AGN. In conclusion, the presence of
24\mic~emission in passive candidates is not totally unexpected and
likely associated with AGN activity.

\subsubsection{Statistical validation of sample} 

Since the available ALMA data is not deep enough to draw any
conclusion for 39\% of the observed candidates, we tried to validate
the passive nature of the entire sample in a statistical sense. To
this aim, we adopted 1$\sigma$ limits in the following.

We first repeated the procedure described above on the stacked flux in
Band 6 and 7, which is compared to the collection of of star-forming
solutions of all sources included in the stacks. We found that the
sample is on average consistent with being passive (a group of
solutions with low SFR values, below the ALMA limits, is observed in
Band 7, but it can be ascribed to a minor fraction of sources,
i.e. 10\% of the stacked sample).  Once again the result is robust
against the uncertainty in modelling the IR SED.  Indeed, only when
adopting two of the models (\citealt{pope08} and \citealt{schreiber18}
with high dust temperature) the results are not conclusive, but again
due to a minority of sources.

\begin{figure}
    \centering
    \includegraphics[width=0.65\columnwidth,angle=90]{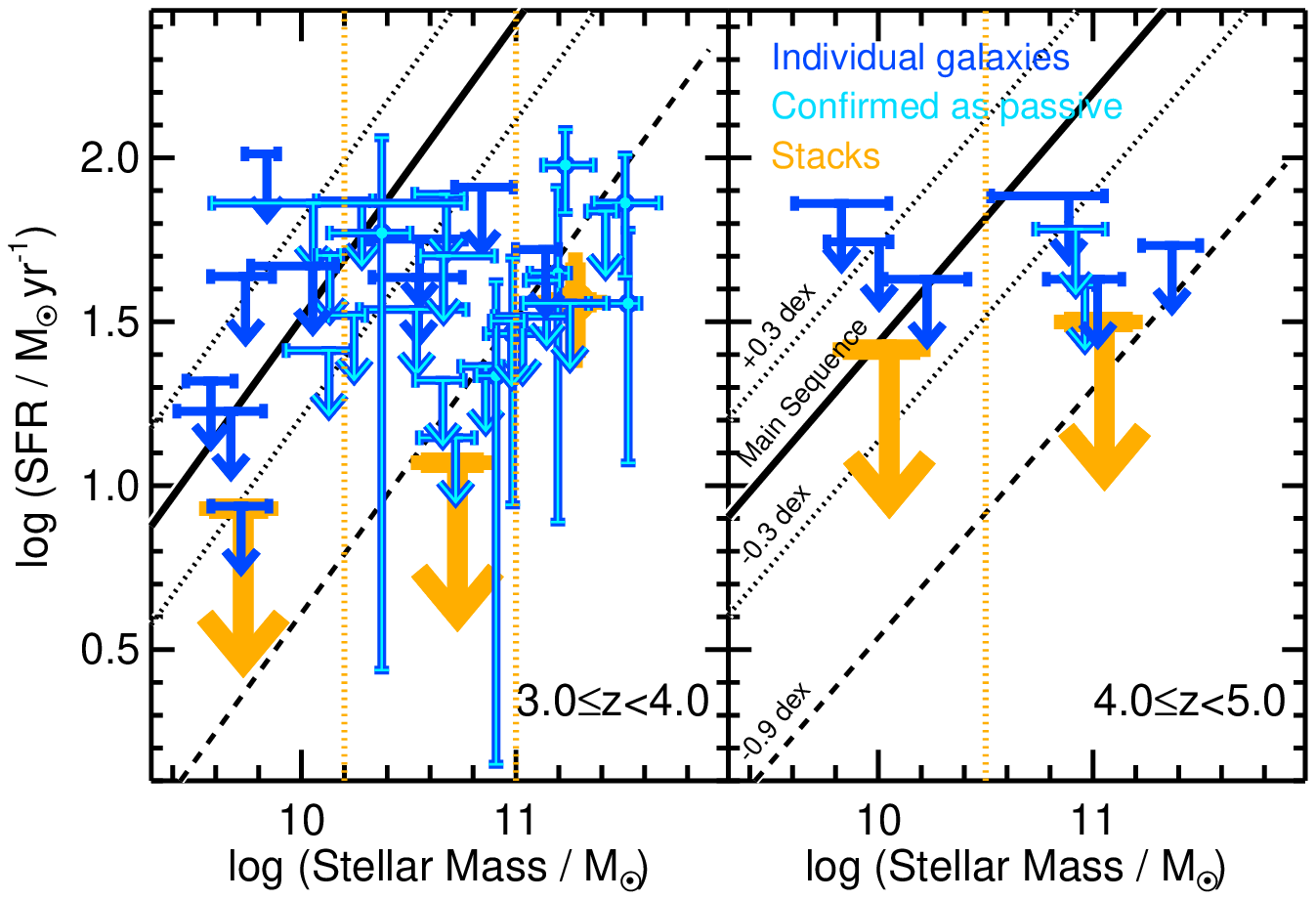}
    \caption[]{Location of the passive candidates observed by ALMA on
      a SFR–-stellar mass diagram, in two redshift bins, based on
      their sub-mm SFR. Arrows represent 1$\sigma$ upper limits on the
      SFR.  1$\sigma$ error bars on the stellar masses have been
      conservatively calculated by setting the redshift as a free
      parameter.  Solid lines show the observed MS of
      \cite{santini17}. Dotted lines are 1$\sigma$ above and below the
      MS (estimated from the observed 0.3 dex scatter), while the
      dashed line is 3$\sigma$ below it. Individual sources are shown
      in blue. Robust candidates (i.e. individually confirmed as
      passive at a $3\sigma$ confidence level) have been overplotted
      in light blue. Large orange symbols show the stacks in bins of
      stellar mass, marked by the orange dotted vertical lines.  }
             \label{fig:ms}
\end{figure}

We then adopted an independent approach to verify how the SFR of the
candidates compares with that of typical star-forming galaxies. We
made use of the stellar mass inferred from the optical fit (a much
more robust parameter than the SFR, as discussed above; see also
\citealt{santini15}) and of the sub-mm-based SFRs, and compared the
position of the candidates on a SFR-stellar mass diagram with the
location of the Main Sequence (MS) of star-forming galaxies. In
Fig.~\ref{fig:ms} we show all candidates observed by ALMA divided into
two redshift bins, compared with the observed (i.e. not corrected for
the Eddington bias) MS inferred from HST Frontier Field data by
\cite{santini17}.  We found that 59\% of the candidates are located
below the 1$\sigma$ scatter of the MS (0.3 dex), while 39\% and 27\%
are at least 2$\sigma$ and 3$\sigma$ below of the MS, respectively,
i.e. in the region where one expects to find sources that have
completely exhausted their star formation.  The fraction of sources at
least 1$\sigma$ / 2$\sigma$ / 3$\sigma$ below the MS is in the range
39-80\% / 24-68\% / 2-41\%, depending on the choice of the IR template
(46-80\% / 34-68\% / 5-41\%, respectively, if the most extreme
template is ignored).  We highlight on the figure the candidates that
have been individually and robustly confirmed
(Sect.~\ref{sec:validation}). The majority of them are located below
the MS. However, a fraction of the confirmed sources have less
stringent constraints, placing them around the MS. We remind that,
with the exception of one source with huge error bars extending down
to the passive area of the diagram, the rest of the candidates around
the MS only have upper limits on the SFR: the knowledge of their exact
position is hampered by the available data, and they may in principle
have much lower levels of SFR. The fact that some of them are
individually confirmed as passive makes us confident that even the
sources with shallow sub-mm constraints preventing their individual
confirmation may actually be intrinsically passive.

We finally stacked sources in bins of stellar mass (orange symbols in
Fig.~\ref{fig:ms}). Once again, the stacking analysis supports the
passive classification of the whole sample. Indeed, the stacked SFR is
well below the MS, with the exception of the lowest mass bin in both
redshift intervals, where available sub-mm data is not deep enough to
reach the lower level of the typical SFR of low-mass galaxies.  The
only IR template unable to confirm the stacking results with high
significance is the high dust temperature template of
\cite{schreiber18}, yielding a high-z/high-M stacked SFR only
$\sim$1.3$\sigma$ below the MS.

We note that, once the sample has been confirmed as passive, either
for individual sources or statistically, ALMA fluxes can be fit with
the IR passive template of \cite{magdis21}. This template yields
$\sim$5 times lower SFR, and places 95\% / 78\% / 66\% of the
candidates 1$\sigma$ / 2$\sigma$ / 3$\sigma$ below the MS (basically
all sources with $\log M/M_\odot>10.4$ are at least 0.9 dex below the
MS according to this template).

\subsection{Conclusions} 

We used ALMA archival observations to confirm the passive nature of
our candidates.  By means of the (lack of) sub-mm emission, we could
validate the passive classification for 61\% of the candidates
observed by ALMA at a 3$\sigma$ confidence level.  At the same time,
based on the available data, we could not reject any of the
candidates, and found that the sample is on average consistent with
being correctly classified as passive. This result is solid against
the large systematics associated with the choice of the IR template
when computing the SFR from sub-mm emission. Only one model, the one
characterized by the highest dust temperature, appears to be unable to
statistically confirm the entire sample with high significance, but we
note that this is the most extreme template, and the assumed
temperature is at the highest envelope of the $T_d$ distribution
(\citealt{faisst20}, see also the extrapolation to high-$z$ of the
results of \citealt{magnelli14} or \citealt{genzel15}).

Once again we confirmed the reliability and robustness of the
photometric selection technique developed in \cite{merlin18} and M19.
Assuming our technique to perform equally well on the sources
unobserved by ALMA, in the following, we consider the entire sample of
101 $3<z<5$ passive galaxy candidates.

\section{Determination of the galaxy Stellar Mass Function (SMF)}\label{sec:mfdet}

To estimate the Stellar Mass Function (SMF) we adopted the stepwise
method, a non-parametric approach that has the advantage of not
assuming an {\it a-priori} shape for the SMF
\citep[e.g.][]{takeuchi00,bouwens08,weigel16}. Instead, the stepwise
estimate assumes that the SMF can be approximated by a binned
distribution, where the number density $\phi_j^{true}$ in each mass
bin $j$ is a free parameter.

In particular, we adopted the procedure presented in
\cite{castellano10b} (Sect. 6.2.2).  We assumed a fixed, constant,
reference density $\phi_{ref}$ over the entire mass range and
exploited a simulation to compute the distribution of observed stellar
masses originating in each mass bin, as a consequence of percolation
of sources across adjacent bins caused by photometric scatter or
failure in the selection technique to isolate simulated passive
galaxies.  In order to take into account field-to-field variations in
depth and selection effects, the simulations were run for each field
separately, scaled to the relevant observed areas, and eventually
summed to obtain the total number densities predicted in each mass
bin.  The intrinsic densities are expressed as
$\phi_j^{true} = w_j \cdot \phi_{ref}$, where $w_j$ are the
multiplicative factors to the reference density of the simulation. The
intrinsic densities, hence the multiplicative factors, and relevant
uncertainties, which best reproduce the observed number densities
$\phi_i^{obs}$ of our survey, were determined by inverting the linear
system $\phi_i^{obs}=\sum_j (S_{ij} \phi_j^{true})$, where $S_{ij}$
are the elements of the transfer function computed from the simulation
(i.e. $S_{ij}$ is essentially the number of galaxies in the mass bin
$i$ scattered from the bin $j$ and taking into account the sources
missed by the selection).  The bin size was chosen as a compromise
between the need to have reasonable statistics in each bin and the
desire to sample the shape of the SMF with good mass resolution.

Simulated masses were obtained by redshifting between $z=3$ and $z=5$
a subset of the synthetic spectra of the stellar libraries. The
library parameters have been chosen in order to reproduce the observed
mass-to-light ratios in the rest-frame $I$ band of the passive
candidates and of the entire $3\leq z < 5$ sample. For the passive
sources we only considered passive templates. Each template has then
been normalized to stellar masses between 9 and 12 in the logarithmic
space. The final simulations are made of 32893 templates for the
``top-hat'' library and 43044 templates for the $\tau$-models library.
The simulations were designed to reproduce both the average and the
scatter of the observed S/N as a function of magnitude to provide the
closest match between observed and simulated data, as done in
\cite{castellano12}. Differences in depth and therefore selection
effects among the different fields are taken into account by running
the simulation independently for each of them. In addition, the two
GOODS fields have inhomogeneous coverage in the HST bands. While the
difference in depth between the CANDELS wide and deep areas is
accounted for by the scatter of the error-magnitude relation, this is
not true for the ultradeep GOODS-S area (4.6 arcmin$^2$,
\citealt{guo13}). For this reason, we treated the latter as a separate
field, with its own simulation.  For each field, synthetic fluxes have
been obtained by convolving the synthetic templates with the filter
response curves and have been perturbed with noise in a way to mimic
the field properties in terms of depth.  We then fitted these
simulated catalogues and selected simulated passive galaxies in the
very same way as done on real data. For the total population, we
simply selected $H<27$ galaxies. We verified that the resulting
mass-to-light ratio distribution is in good agrement with the input
one (this makes us confident that mass-to-light ratios are not
substantially modified by observational errors).

Although the simulation covers stellar masses from $10^9$ to
$10^{12} M_\odot$, in the following we will consider the SMF on a
narrower range of masses.  Indeed, the required corrections below
$10^{10} M_\odot$ make the SMF very uncertain at these low
masses. However, low mass bins need to be included in the procedure to
consider the effect of the Malmquist bias, i.e. to account for low
mass sources scattered to larger measured masses.  The overall
corrections applied to the observed counts are within a factor of 3,
with the only exception of the lowest mass bin at $z>3.5$, affected by
a higher level of incompleteness, where the required correction is a
factor of 5 and 10, respectively, at intermediate and high redshift.

The stepwise method has the advantage of taking into account mass
completeness issues, incompleteness in the passive selection,
photometric errors (i.e. the percolation of sources across adjacent
mass bins), and the Eddington bias.  The latter is naturally accounted
for without any a-posteriori simulation.  As shown by
\cite{davidzon17}, stepwise results show very good agreement with the
standard $1/V_{max}$ method. As an additional test, we verified that
the SMF computed with the stepwise approach on the GOODS-S sample,
using the very same photometric catalogue, redshifts and sample
selections of \cite{grazian15}, nicely agrees with their $1/V_{max}$
points.

The cumulative area over which the SMF was calculated is 969.7
arcmin$^2$.  Cosmic variance has been added using the QUICKCV code of
\cite{moster11}. It computes relative cosmic variance errors as a
function of stellar mass, in bins very similar to the ones used by us,
up to $10^{11.5}M_\odot$. For the largest mass bin, we adopted the
value for the 11-11.5 logarithmic bin incremented by 50\%. We verified
that this particular choice does not significantly affect our results.
Since cosmic variance not only depends on the total area surveyed but
also on the survey geometry, following \cite{driver10}, we reduced
these relative errors by $\sqrt{N}$, where $N=5$ is the number of
non-contiguous fields of similar area.

Finally, Schechter functions have been fitted to the stepwise points
and uncertainties.

As discussed in Sect.~\ref{sec:passivesel}, we did not exclude AGN
candidates from the sample.  Anyhow, we verified that the exclusion of
the 9 X-ray detected candidates among the passive sample (and similar
exclusion from the total sample) does not change our results.

\section{The SMF and the Stellar Mass Density of $3\leq z<5$ passive galaxies}\label{sec:mf}

\begin{figure*}[!t]
    \centering
   \includegraphics[width=0.9\columnwidth,angle=270]{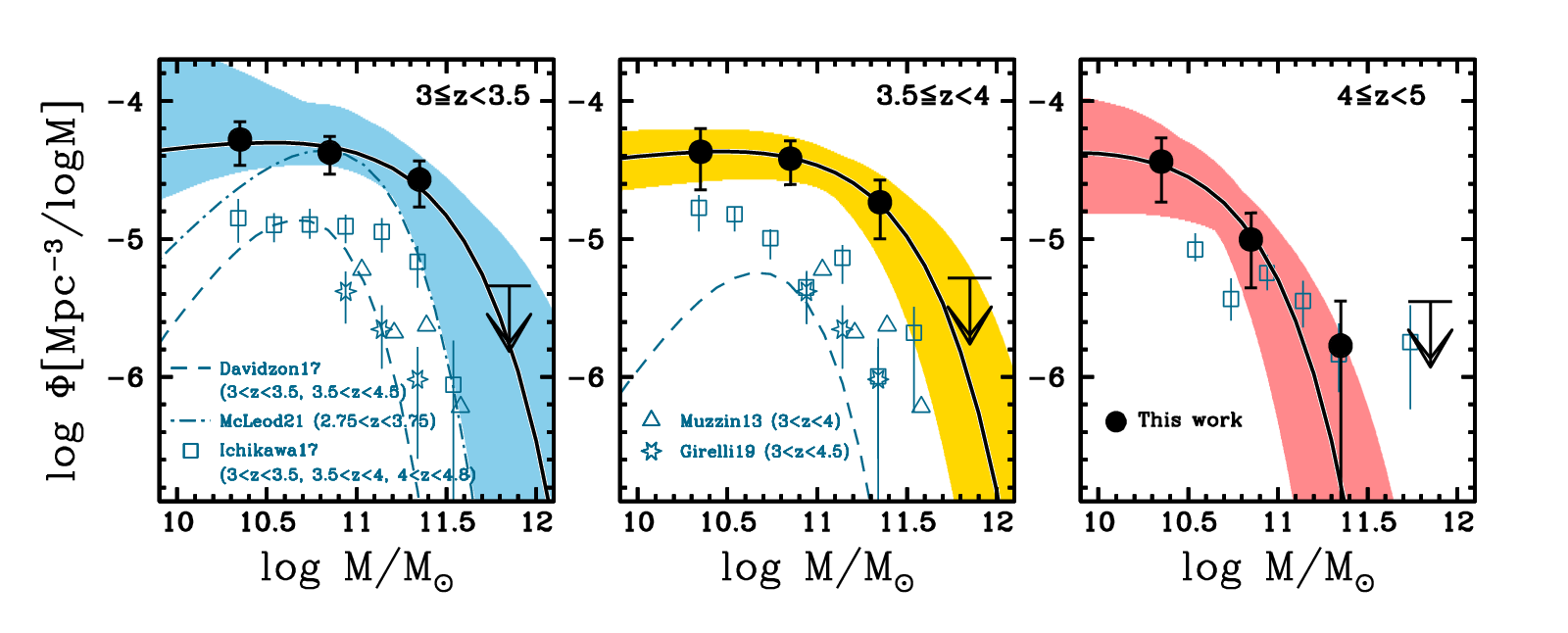}
   \caption[]{SMF for passive galaxies in the five CANDELS
     fields. Black solid points show the stepwise results, with error
     bars accounting for Poissonian and cosmic variance
     uncertainties. Black curves show the Schechter fits and the
     coloured shaded areas indicate the regions at 68\% confidence
     level considering the joint probability distribution function of
     the Schechter parameters.  Blue open symbols and lines show the
     SMF for passive galaxies taken from the literature, scaled to a
     Salpeter IMF, in the same (or very similar) redshift bins
     (\citealt{ichikawa17} as squares, \citealt{davidzon17} as dashed
     line and \citealt{mcleod21} as dot-dashed line), and at slightly
     different redshifts (\citealt{muzzin13} as triangles and
     \citealt{girelli19} as 6-points stars).  }
              \label{fig:mfpas}
\end{figure*}

The resulting SMF for the passive population, computed in three
different redshift bins, is shown in Fig.~\ref{fig:mfpas} and reported
in Table~\ref{tab:mf}. We also show the best-fit Schechter functions
and the associated 68\% probability confidence regions.  Due to the
lack of constraints at low masses and larger uncertainties at high
redshift, at $z>3.5$ we fixed the value of slope $\alpha$ to its best
fit value in the lowest redshift bin. The best fit Schechter
parameters are listed in Table~\ref{tab:bestfit}.

\begin{figure}
  \includegraphics[width=\columnwidth]{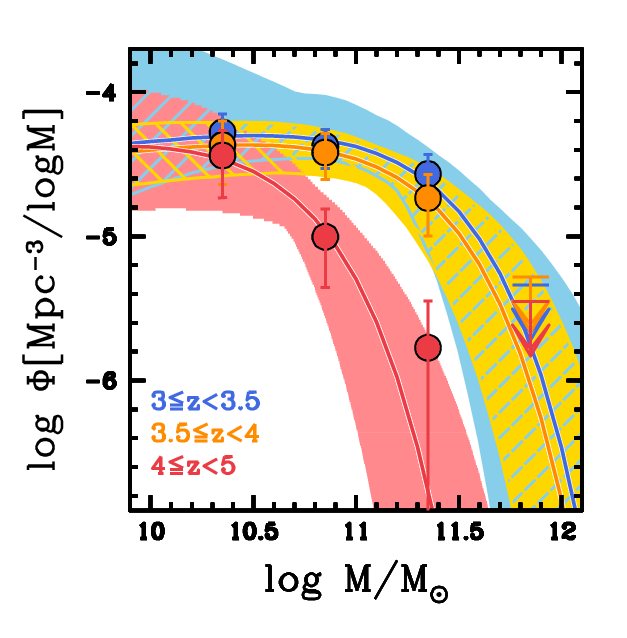}
    \caption{Evolution in the SMF of passive galaxies from $z=5$ to
      $z=3$.  Points, lines and shaded regions are as in
      Fig.~\ref{fig:mfpas}.}
              \label{fig:mfevol}
\end{figure}

A strong evolution in the ``knee'' of the SMF ($M^*$) is observed
around $z\sim4$.  For a better visualization of the evolution in the
SMF we show in Fig.~\ref{fig:mfevol} the three redshift bins
simultaneously.  Despite the large associated uncertainties, we find a
clear evolution in the passive galaxy population from $z=5$ to $z=3$.
While both the stepwise points and the best-fit Schechter functions
are very similar in the two lowest redshift bins, we observe a
pronounced evolution beyond $z\sim 4$, indicating that we are
witnessing the epoch of {\it passivization} of massive galaxies. While
the lowest mass passive galaxies seem to be already in place at these
redshits, the largest structures have not had the time to form yet at
$z>4$.  We note that this result is independent on the choice of the
mass binning.

\begin{figure*}
    \centering
  \includegraphics[width=1.1\columnwidth,angle=270]{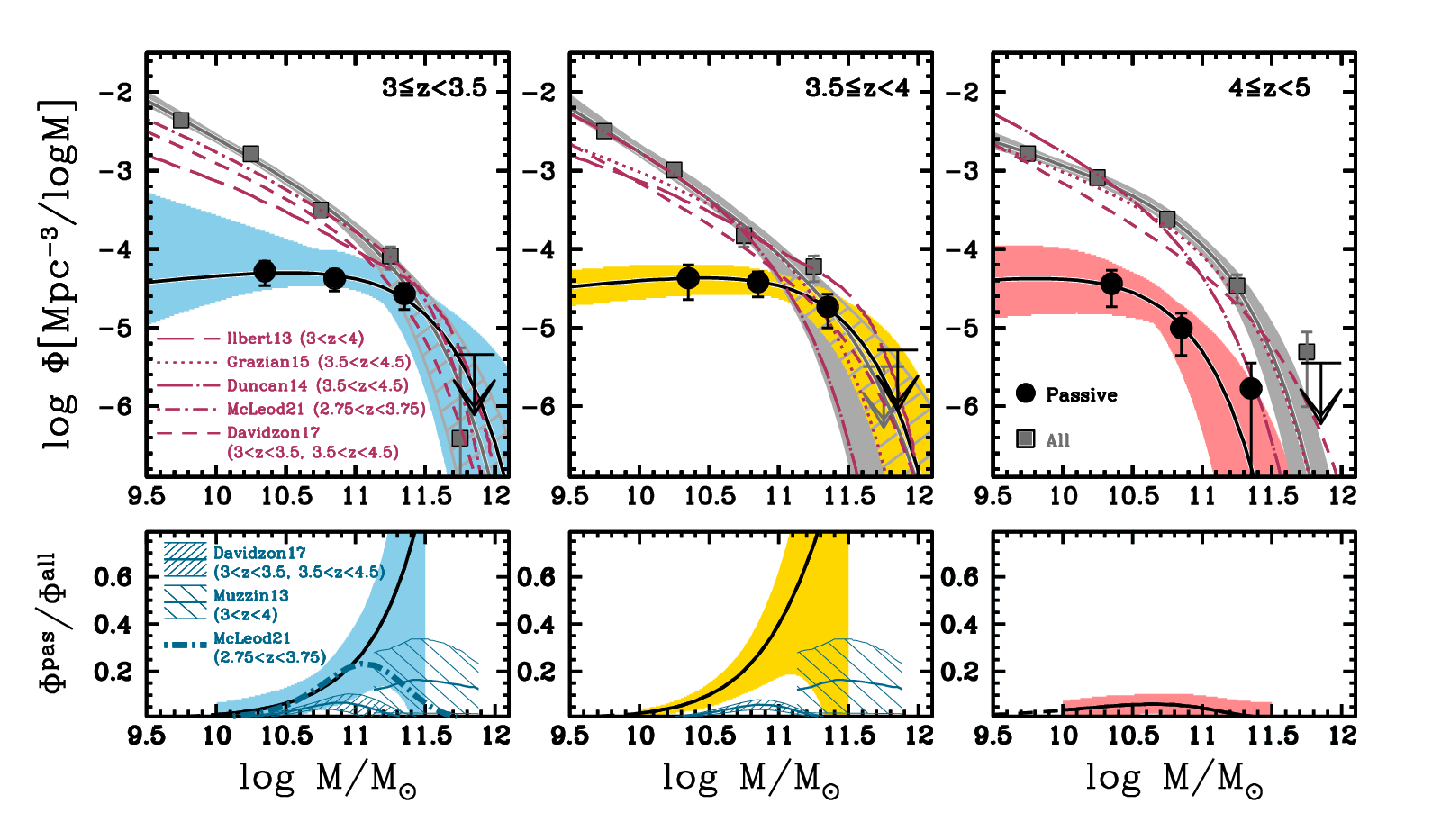}
  \caption[]{{\it Upper panels}: SMF of passive candidates (black
    circles and curves, coloured shaded regions) vs all $H<27$
    galaxies (gray squares and curves, gray shaded regions). Shaded
    regions indicate the regions at 68\% confidence level considering
    the joint probability distribution function of the Schechter
    parameters.  Dark red curves show SMF from the literature in
    similar, but not identical, redshift bins (as indicated in the
    legend), all scaled to the same IMF: \cite{ilbert13} (long
    dashed), \cite{duncan14} (dot-long dashed), \cite{grazian15}
    (dotted), \cite{davidzon17} (dashed) and \cite{mcleod21}
    (dot-dashed). {\it Lower panels}: Ratio of the passive and global
    Schechter functions. The shaded area and solid curve show the mass
    range where the comparison is meaningful (i.e. not affected by
    paucity of galaxies nor extrapolated due to the lack of
    observational data).  We also show the passive fractions published
    by \cite{davidzon17} (narrow-spaced shaded area) and
    \cite{muzzin13} (wide-spaced shaded area), and the ratio between
    the passive and the total SMF of \cite{mcleod21} (dot-dashed
    line).  }
              \label{fig:mfpasallratio}
\end{figure*}

With the aim of understanding the relative abundance of passive
galaxies, we computed the SMF for the entire population. The results
are shown in Fig.~\ref{fig:mfpasallratio} (gray stepwise points and
gray Schechter fits).  We also show several SMF for the entire galaxy
population taken from the literature
\citep{ilbert13,duncan14,grazian15,davidzon17,mcleod21}. Different
fields, photometric quality, sample selections, methodology and
corrections for incompleteness are responsible for the large spread in
the SMF observed at such high redshifts. For this reason, for a fair
comparison with the passive population, instead of using a reference
SMF from the literature we decided to recompute the SMF for all
CANDELS $H<27$ galaxies using the very same method applied to the
passive candidates.  We note that our estimates tend to be on the
upper envelope of the distribution of the various results from the
literature. This is a consequence of our choice to include AGNs in the
sample.

\begin{table}
\centering
\begin{tabular} {cccc}
\hline \hline 
\noalign{\smallskip} 
 $\log M/M_\odot$ & \multicolumn{3}{c}{$\phi$ [$10^{-5}$/Mpc$^3$/dex]}\\
\noalign{\smallskip} 
 & $3<z<3.5$ & $3.5<z<4$ & $4<z<5$\\
\noalign{\smallskip} 
\hline 
\noalign{\smallskip} 
      10.35 &        5.25 $\pm$  1.76 &   4.29 $\pm$ 2.00 &  3.64 $\pm$   1.74 \\         
      10.85 &        4.24 $\pm$  1.17 &   3.82 $\pm$ 1.19 & 0.99 $\pm$ 0.53 \\            
      11.35 &        2.70 $\pm$  0.84 &   1.84 $\pm$ 0.71 & 0.17 $\pm$ 0.18 \\            
      11.85 &     $<$  0.49 &             $<$ 0.53 &        $<$       0.36 \\             
\noalign{\smallskip} \hline \noalign{\smallskip}
\end{tabular}
\caption{Stellar Mass Function of passive galaxies as inferred from a
  stepwise method.  1$\sigma$ uncertainties include Poissonian errors and cosmic variance.
}\label{tab:mf}
\end{table}

While at $z<4$ the massive tail of the SMF of passive and all galaxies
are rather similar, they diverge below $\sim 10^{11}M_\odot$.  The
lower panels of Fig.~\ref{fig:mfpasallratio} show the ratio between
the passive and the global SMF ($\phi^{pas}/\phi^{tot}$).  As
expected, we observe a trend with stellar mass, with passive galaxies
being more abundant at high masses relative to the global population.
While making a negligible fraction of the total population at
$\sim10^{10}M_\odot$, $z<4$ passive galaxies become dominant at large
stellar masses. Although the associated uncertainty is too large to
estimate a precise fraction, passive galaxies could make up to
$\sim$30\% of the total galaxy population at $M\sim10^{11}M_\odot$ and
more than 50\% at larger masses.  We note, however, that while the
passive SMF is nicely fitted by a Schechter function at stellar masses
between $10^{10}$ and $10^{11.5} M_\odot$, its shape is poorly
constrained above this value. Due to the paucity of such sources,
larger surveys are needed to better constrain the high-mass tail of
the passive SMF. Consistently with the pronounced evolution around
$z\sim 4$, the fraction of passive galaxies earlier than this epoch is
very low. Even at the highest masses, passive galaxies constitute no
more than a few percent of the total galaxy population at $z>4$.

\begin{figure*}
\centering
   \includegraphics[width=1.3\columnwidth,angle=270]{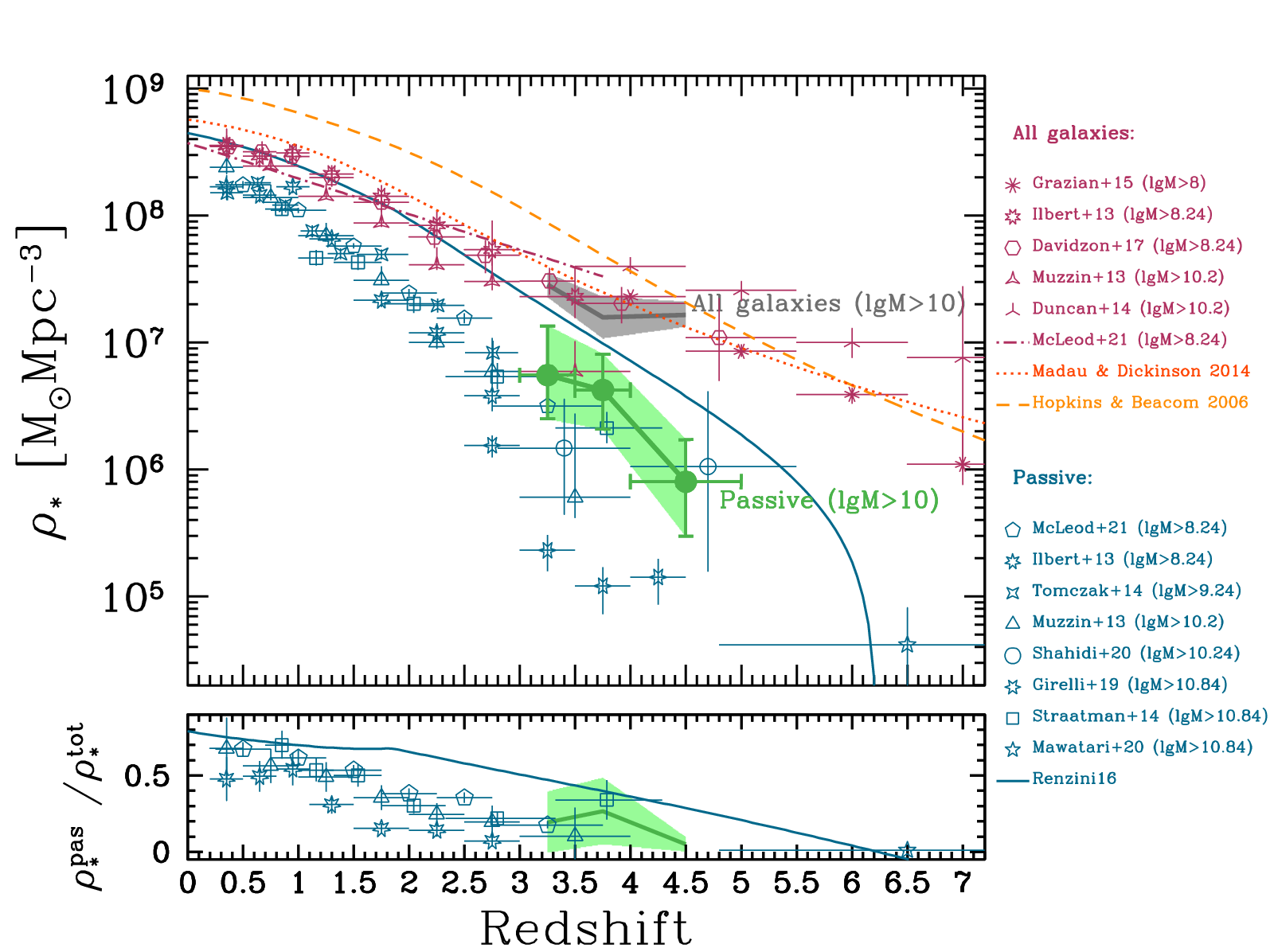}
   \caption{{\it Upper panel}: Evolution in the Stellar Mass Density
     of passive galaxies above $10^{10} M_\odot$ (green symbols and
     shaded area). As for comparison, we also plot the SMD of all
     $H<27$ galaxies integrated above the same mass limit (gray shaded
     region). In both cases, the shaded regions represent the
     1$\sigma$ uncertainty. Results are compared to a number of
     literature works, as listed in the legend, both for passive
     galaxies (blue open symbols) and for the entire population (dark
     red open symbols and dot-dashed line). Literature estimates and
     their mass integration limits have been scaled to a Salpeter IMF,
     when needed. The dashed orange and dotted red lines are the SMD
     derived by integrating Star Formation Rate Density from
     \cite{hopkins06} and \cite{madau14}, respectively, assuming a
     constant recycle fraction of 28\%. The solid blue line shows the
     prediction for the SMD in passive galaxies by \cite{renzini16}.
     {\it Lower panel}: Ratio between the mass density of passive
     galaxies and of the overall galaxy population above
     $10^{10} M_\odot$ (green solid line and solid region). We also
     report the passive fractions in terms of mass densities measured
     by \cite{muzzin13} (triangles), \cite{straatman14} (squares),
     \cite{ilbert13} (7-points stars), \cite{mcleod21} (pentagons),
     \cite{mawatari20} (5-points star) and predicted by
     \cite{renzini16} (solid line).}
              \label{fig:md}
\end{figure*}

We finally estimated the Stellar Mass Density (SMD) accounted for by
passive galaxies at different epochs by integrating the best-fit
Schechter functions reproducing their SMF. In order not to be affected
by large extrapolations at low stellar masses, we integrated the SMF
from $10^{10} M_\odot$ to $10^{13} M_\odot$. The results and
associated 1$\sigma$ uncertainties are shown in Fig.~\ref{fig:md} as
green points and shaded area (while the gray shaded region represents
the global population).

An increase in the SMD of passive galaxies by a factor of 7 from
$z\sim 5$ to $z\sim 3$ is observed.  In the lower panel of
Fig.~\ref{fig:md} we show the passive fraction in terms of stellar
mass density, i.e. the ratio between the passive and the total SMD
($\rho^{pas}/\rho^{tot}$). 20-25\% of the total mass assembled by
redshift $\sim$3 is accounted for by passive galaxies, while this
fraction reduces to $\sim$5\% at $z>4$.

We verified that our results for the abundance of passive galaxies, in
terms of their contribution to both the SMF and the SMD, remain
unchanged when AGNs are excluded from the sample.  They may instead be
affected by the inclusion of HST-dark galaxies, $z\gtrsim 3$ red
galaxies that are undetected on the $H$ band image
\citep{huang11,caputi12,wang16,wang19,alcaldepampliega19},
whose 
abundance relative to the underlying $H$-detected red massive galaxy
population increases with redshift and stellar mass.  These objects
are not included in our sample because of the $H<27$ cut. They are
taken into account by our simulation to some extent that is
impossibile to quantify exactly, because of the difficulty of
mimicking their observed mass-to-light ratio distribution, being them
not contained in the CANDELS catalogs.  We can estimate an upper limit
to their additional contribution to the overall statistics of high-$z$
quiescent galaxies by correcting our passive and global SMF according
to the results of \cite{alcaldepampliega19}. They estimated the
contribution of HST-dark galaxies to a mass-selected sample in bins of
redshift and stellar mass, and estimated that 30\% of them are
quiescent or post-starburst galaxies.  While the effect of including
HST-dark would be negligible at $z<4$, at the highest redshifts the
passive and global SMFs would increase by $\sim$0.2-0.3 ($\sim$1) dex
and $\sim$0.05 ($\sim$0.1) dex, respectively, at intermediate (high)
masses. This would however leave our results on the abundance of
passive galaxies relative to the global population almost unaffected,
due to the already very low fraction of passive galaxies at $z>4$.  We
note, anyway, that this is a conservative evaluation of the
contribution of HST-dark galaxies to the passive fraction, because of
the less stringent definition of quiescent/post-starburst galaxy of
\cite{alcaldepampliega19} compared to the one adopted by us.

Overall, we can conclude that we, despite the difficulties associated
with the study of high-$z$ passive galaxies, these sources are already
in place in the young Universe. They make up a significant fraction
($>$50\%) of massive ($M\gtrsim 10^{11}$-$10^{11.5}M_\odot$) galaxies
up to $z=4$, and account for $\sim$20\% of the total stellar mass
formed by that time. On the contrary, only a few percent passive
galaxies are observed earlier than this epoch. We are witnessing the
emergence of the quiescent population, i.e. their {\it passivization},
in the redshift interval spanned by this work.

\begin{table*}
\centering
\begin{tabular} {cccccc} 
\hline \hline 
\noalign{\smallskip} 
Redshift & N & $\alpha$ & $\log M^*/M_\odot$ & $\log \phi_*$/Mpc$^{-3}$ & $\log \rho_{>10}/(M_\odot$Mpc$^{-3}$)\\
\noalign{\smallskip} \hline \noalign{\smallskip}
3.0 - 3.5 & 58 & -0.81 $\pm$ 0.52 & 11.23 $\pm$ 0.38 & -4.44 $\pm$  0.35 & 6.74$^{+0.39}_{-0.34}$ \\
3.5 - 4.0 & 31 & -0.81 & 11.19 $\pm$ 0.20 & -4.51 $\pm$ 0.14 & 6.62$^{+0.28}_{-0.31}$ \\
  4.0 - 5.0 & 12 & -0.81 & 10.55 $\pm$ 0.22 & -4.52 $\pm$ 0.30 & 5.90$^{+0.33}_{-0.43}$ \\
\noalign{\smallskip} \hline \noalign{\smallskip}
\end{tabular}
\caption{Best-fit parameters and their 1$\sigma$ uncertainties in the different redshift
  intervals derived from fitting the stepwise SMF with a Schechter function.  At $z>3.5$, the Schechter slope $\alpha$ has been fixed to the value in the lowest redshift bin.  The second column indicates the numbers of galaxies in each redshift bin based on which the SMF is actually computed. 
  The last column reports the corresponding mass density $\rho$
  obtained by integrating the SMF from $10^{10} M_\odot$ to  $10^{13} M_\odot$. 
}\label{tab:bestfit}
\end{table*}

\section{Comparison with the literature}\label{sec:comparison}

\subsection{Comparison with previous observations}

At variance with other aspects of galaxy evolution over which a large
consensus exists by now, the study of high-$z$ passive galaxies, and
in particular the details such as their abundance, is still matter of
debate, and current surveys have not found a final and joint solution
yet. Early passive galaxies are rare sources, and several differences
among the various analyses may be responsible for the variance among
their results. Firstly, the photometric quality is crucial to detect
these faint and elusive sources, and different filter combinations and
depths affect the selection in a non straightforward way. Secondly,
given the paucity of these sources, field-to-field variations may play
an important role. Thirdly, different selection techniques may lead to
different candidates, as discussed at length in M19. Finally, the
methodology adopted to compute the SMF and the correction for
observational incompleteness are essential ingredients in the
analysis, as discussed in the following.

We compare here our findings with the few previous works that analyse
the SMF of passive galaxies at similar redshifts. We plot on
Fig.~\ref{fig:mfpas} the SMF of \cite{davidzon17} and
\cite{ichikawa17}, using the same redshift bins adopted by us, and the
estimate of \cite{mcleod21}, covering a redshift space slightly larger
than our first interval but equally centred at $z\sim 3.25$. As for
reference, we also plot the results of \cite{muzzin13} and
\cite{girelli19}, although their SMF are estimated on redshift
intervals not coincident with ours, making the comparison tricky.  For
this reason, we will not discuss their results in the following.  As
for \cite{ichikawa17} results, we sum their estimate for passive and
recently quenched galaxies. For \cite{davidzon17} and \cite{mcleod21}
we plot their Schechter fits, which include the correction for the
Eddington bias. All these works selected passive galaxies through a
colour selection (based on either observed or rest-frame colours). All
of them, except \cite{mcleod21} results, are based on the shallower
COSMOS field.

Our SMF of passive galaxies at $z\sim 3.25$ nicely agrees with that of
\cite{mcleod21} at intermediate mass, and is marginally 1$\sigma$
consistent with it at the high mass end. This mismatch may be due to
fluctuations in the relatively small area probed by us. Overall,
although we find larger densities, given the different parent sample,
sample selection and methodology, the consistency is more than
satisfactory.  The SMF measured by us at $z\sim 3.25$ and $z\sim3.75$
are higher than those of \cite{davidzon17} by a factor of $\sim$4 to
$\sim$8 at the peak and by more than a decade at larger masses. At
$z<4$ the mismatch with \cite{ichikawa17} is a factor of $\sim$4 at
the peak, while the massive tail is marginally consistent with our
results. The agreement with the latter study improves at $z>4$, where
the two estimates are well consistent within the error bars.

In the lower panels Fig.~\ref{fig:mfpasallratio} we compare the
fraction of passive sources as a function of stellar mass
($\phi^{pas}/\phi^{tot}$) with previous works, less prone to the
systematics associated with the various SMF. Once again, our results
nicely agree with those of \cite{mcleod21} at $M<10^{11}M_\odot$, but
diverge at higher masses (though remain consistent within the error
bars).  They are also consistent within the uncertainties with those
of \cite{davidzon17}, while we observe higher fractions than
\cite{muzzin13}.

Literature results on the SMD of passive galaxies show a very large
variance, spanning more than an order of magnitude at the redshifts
probed by this work. Our results tend to be on the upper envelope of
the ranges spanned by previous works, and are consistent with the
analyses of \cite{mcleod21}, \cite{straatman14} and \cite{shahidi20}
(Fig.~\ref{fig:md}). We note however that the integration limits are
not always consistent. For example, \cite{mcleod21} integrates down to
a lower mass limit, but their lower SMD is likely explained by their
steeper SMF at both low (mostly) and high stellar masses.

The integrated passive fraction (i.e. the ratio between the SMD of
passive vs all galaxies, $\rho_*^{pas}/\rho_*^{tot}$, shown in the
lower panel of Fig.~\ref{fig:md}) is consistent with that reported by
\cite{muzzin13} and \cite{mcleod21} at $z\sim 3-3.5$.
\cite{straatman14} found a higher ratio of 34\%, still consistent with
our estimate, with the difference likely explained by their higher
integration limit (indeed, passive galaxies are more abundant at large
stellar masses, as shown in Fig.~\ref{fig:mfpasallratio}). However,
their higher value might also reflect their higher number density and
could be explained by their shallower selection criterion (see
discussion in M19).  The constant decline in the quiescent fraction
observed from the local Universe to high redshift could flatten at
$z\gtrsim2-3$, as suggested by \cite{straatman14}, all the way to
$z\sim4$, before dropping to zero at $z>5$ \citep{mawatari20}.

The fact that we measure higher SMF is somewhat expected thanks to the
overall better CANDELS photometric quality (compared to the analyses
based on COSMOS) and to the selection technique adopted to single out
passive sources. Our selection does indeed conservatively reject non
secure sources, but at the same time it is able to identify candidates
that are missed by colour selections at these redshifts (see
\citealt{merlin18} and M19 for an extensive discussion; see also
\citealt{schreiber18c}, \citealt{deshmukh18} and \citealt{carnall20}
regarding the incompleteness of the UVJ colour selection).

A trend of selecting a larger number of passive sources was already
observed in the number density by M19. The mismatch is however larger
here, and larger than reported by other authors \citep{shahidi20},
because of the corrections applied for incompleteness by our
method. While the majority of previous works (except
\citealt{mcleod21}) only correct the observed mass distribution for
incompleteness in stellar mass due to the survey depth, our procedure
also takes into account the intrinsic incompleteness in the passive
selection. Indeed, as explained in Sect.~\ref{sec:mfdet}, we applied
to the simulated galaxies the very same selection technique adopted to
single out the passive candidates. In this way, our procedure not only
accounts for galaxies with light-to-mass ratio too low to be detected
by our survey, but also for those passive galaxies that are missed by
the selection (i.e. not classified as passive) due to photometric
noise and survey properties (in terms of relative depth of the various
bands).

To assess the relative contribution of observed number densities of
passive galaxies and incompleteness corrections to the mismatch with
respect to previous works, we compared literature results to our
simple mass counts per unit volume, without applying any
correction. The mismatch reduces by a factor of $\sim$2-3: our
observed SMFs are still above the results of \cite{davidzon17}; they
are consistent or below the SMF the \cite{mcleod21}, as expected,
since they also correct for incompleteness; finally, compared to the
results of \cite{ichikawa17}, they are consistent at the highest
redshift and only 1-2$\sigma$ above at $z<4$.  These findings are
expected from the mismatch in the mere number of candidates found by
\cite{davidzon17} and \cite{ichikawa17}, compared to our sample,
scaled by the relative areas.  Indeed, the passive selection of
\cite{davidzon17} and \cite{ichikawa17} encompasses $\sim$8$\times$
and $\sim$3.5$\times$, respectively, less candidates than ours
($\sim$1.4$\times$ less at the highest redshifts,
\citealt{ichikawa17}).  We also compared the uncorrected SMD of
passive galaxies with literature results, and found that it is in line
with the bulk of previous estimates.  Overall, our higher SMF and SMD
derive from a combination of a generally larger number of passive
candidates and a more accurate correction for mass incompleteness,
Eddington bias and incompleteness in the passive selection.

\begin{figure*}
    \centering
  \includegraphics[width=1.1\columnwidth,angle=270]{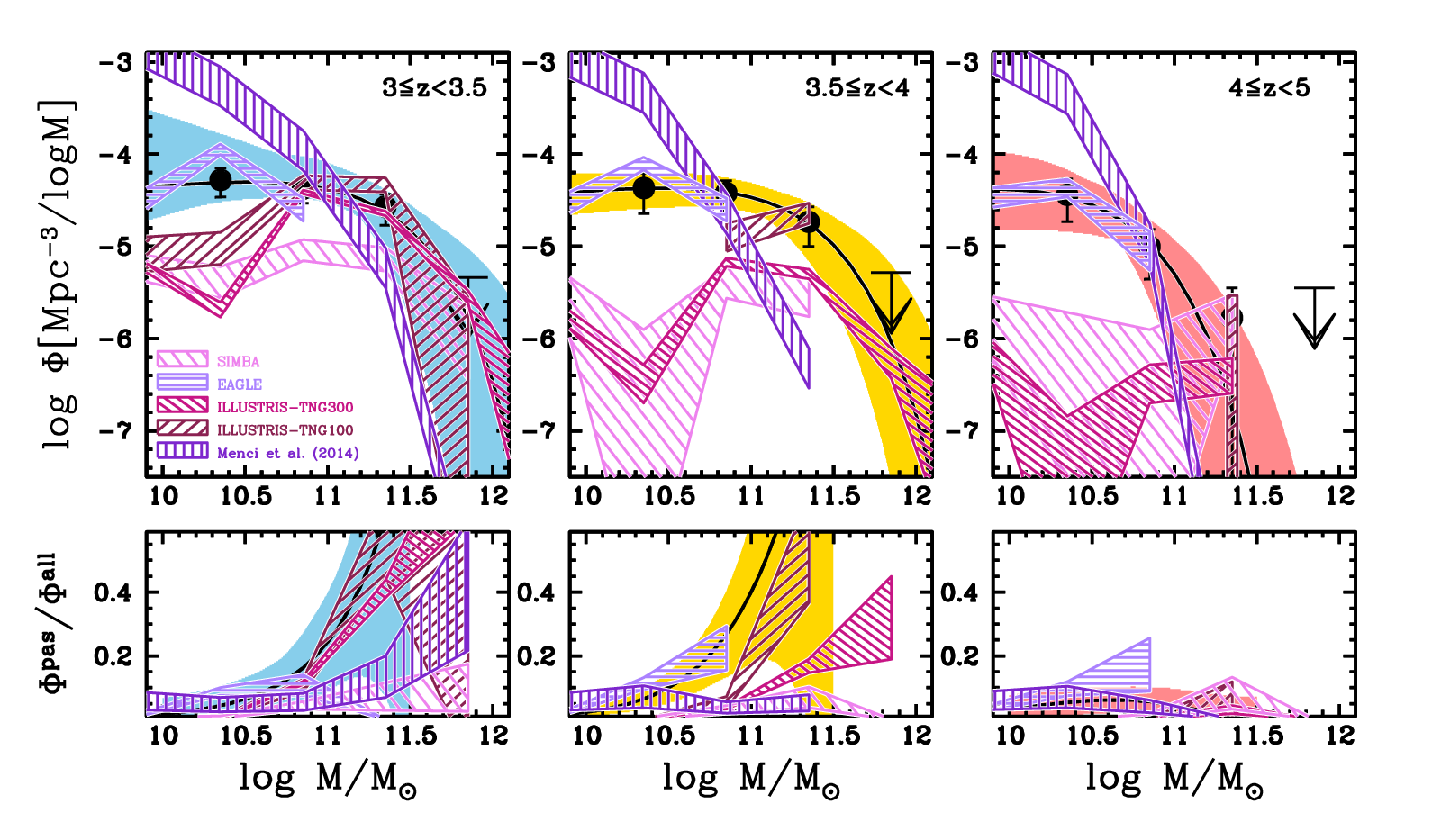}
  \caption[]{{\it Upper panels}: Observed SMF for passive galaxies
    (black points and curves, coloured solid shaded regions) compared
    to the theoretical predictions of the semi-analytic model of
    \cite{menci14} and the hydrodynamic simulations ILLUSTRIS-TNG100,
    ILLUSTRIS-TNG300, EAGLE and SIMBA (see legend), scaled to a
    Salpeter IMF. The bins where the number of model passive galaxies
    is null are not shown. {\it Lower panels}: Ratio of the passive
    and global SMF. The coloured solid shaded area and solid curves
    show the observed fraction, as in
    Fig.~\ref{fig:mfpasallratio}. The open shaded regions show model
    predictions, according to the legend.}
              \label{fig:mfmodel}
\end{figure*}

At variance with the observations just discussed are the predictions
of \cite{renzini16}. He estimated the SMD in passive and star-forming
galaxies by combining the parameterization for the evolution in the
Star Formation Rate Density from \cite{madau14} with that of the sSFR
(=SFR/M) for near MS galaxies of \cite{peng10}, assuming a unitary
slope for the MS. The obtained SMD for the passive population is a
factor of 2 to 5 above our measurements. His prediction for the
passive fraction, shown in the lower panel of Fig.~\ref{fig:md}, is
also higher than our observational data by a similar factor. Indeed,
in his empirical description, passive galaxies start dominating over
the star-forming population at $z>3$.  We note, however, that these
predictions are somewhat indirect, and rely on the assumptions
mentioned above and on the consistency between measurements of SFR and
stellar mass, as discussed in the paper. Moreover, the definition of
passive galaxies is not entirely consistent with ours, as
\cite{renzini16} defines as passive all sources with
sSFR$\ll$sSFR$_{MS}$, including parts of star-forming galaxies such as
quenched bulges and stellar halos, so a higher abundance of passive
sources is expected. On the other side, his method has the advantage
of not being affected by incompleteness in the selection of passive
candidates.

\subsection{Comparison with theoretical predictions}

To put our results in a broader context and understand how they fit
the current theoretical scenario, we compared them with the
predictions of one semi-analytic model (the PANDA model,
\citealt{menci14}) and four hydrodynamic simulations (ILLUSTRIS-TNG100
and -TNG300, \citealt{pillepich18a}, \citealt{nelson19}, EAGLE,
\citealt{schaye15} and SIMBA \citealt{dave19}).

\subsubsection{The models}

The model of \cite{menci14} is an improved version of the
semi-analytic model of \cite{menci06,menci08}. It connects, within a
cosmological framework, the baryonic processes (gas cooling, star
formation, supernova feedback) to the merging histories of the dark
matter haloes, computed by means of a Monte Carlo simulation.  AGN
activity is triggered by galaxy interactions and, in this latest
version, by disk instability. The description of AGN feedback has been
updated by implementing the 2-D modelling for the expansion of
AGN-driven shocks \citep{menci19}.  This model adopts a Salpeter IMF.

ILLUSTRIS-TNG100 and -TNG300 exploited the moving mesh code AREPO
\citep{springel10} and simulated volumes of 100 and 300 co-moving Mpc
on each side ($110.7^3$ and $302.6^3$ Mpc$^3$) with baryonic mass
resolution of $1.4 \times 10^6$ and $1.1 \times 10^7 M_\odot$,
respectively.  EAGLE is a set of simulations based on an updated
version of the N-body Tree-PM smoothed particle hydrodynamics code
GADGET-3 \citep{springel05}. We used the run EagleRefL0100N1504, that
simulated a volume of 100$^3$ Mpc$^3$ with baryonic mass resolution of
$1.81 \times 10^6 M_\odot$.  The SIMBA simulations utilize the Gizmo
cosmological gravity plus hydrodynamics solver
\citep{hopkins15,hopkins17}, in its Meshless Finite Mass (MFM)
version. It has a volume of (100/$h$)$^3$ Mpc$^3$, where $h=0.68$,
with baryonic mass resolution of $1.82 \times 10^7 M_\odot$.  All
these models include baryonic sub-grid physics to simulate star and
black hole formation, stellar and AGN feedback, and metal
enrichment. They assume Chabrier IMF, and have been scaled to a
Salpeter one by multiplying the stellar masses by 1.74
\citep{salimbeni09a}.

\subsubsection{The selection of model passive galaxies}

Accurately simulating the entire observational selection with a
forward modelling approach is extremely time consuming and will be
addressed in a future work (Fortuni et al. in prep.). We therefore
here adopted a standard criterion to select model passive galaxies.
One possibility would be to select passive galaxies based on their
specific SFR, that is required to be lower than a given
threshold. While a fixed cut of $10^{-11}$~yr$^{-1}$ is widely
adopted, it may be too conservative if applied to high redshift
galaxies, as it does not take into account the evolution of the
typical star formation rate. For this reason, we adopted the
time-evolving criterion used by \cite{pacifici16},
\cite{carnall19,carnall20}, as well as applied by \cite{shahidi20} on
theoretical models, i.e. sSFR$<0.2/t_U$, where $t_U$ is the age of the
Universe at the galaxy redshift.

The ILLUSTRIS-TNG and EAGLE simulations release multiple estimates of
stellar mass and SFRs, based on the physical region within which they
are computed.  We adopted the very same stellar masses and SFR as in
our previous work (M19): for ILLUSTRIS-TNG they are defined within
twice the half mass radius, best reproducing the observations (see
M19, \citealt{valentino20} and \citealt{shahidi20}); for the EAGLE
model we considered the aperture which is closest to
$4 \times R_{1/2,100}$, where $R_{1/2,100}$ is the half-mass radius
computed within 100 kpc. For SIMBA we adopted the stellar masses and
SFRs associated with all particles gravitationally bound to the dark
matter halo, as these are the only available in the catalog.  We note
that the aperture choice over which stellar masses and SFRs are
computed may have a significant effect on the selection of model
passive galaxies, especially at $2<z<3$ and $M>10^{10.5}M_\odot$, as
shown by \cite{donnari20}.  In the future, we plan to extend their
work to higher redshifts, and analyze the impact of different
definitions of quenched galaxies, apertures and averaging timescales
for the SFR (Fortuni et al. in prep.).

\subsubsection{The predicted SMF}

We show the comparison of theoretical predictions for the SMF with
observational results in Fig.~\ref{fig:mfmodel}.  Error bars
associated with the hydrodynamic simulations are based on Poissonian
errors. In the semi-analytic model, the uncertainty is caused by the
limited number of merger tree realizations. We computed the relative
error as the Poissonian error on the number $N$ of merger tree
realizations, i.e. $\sqrt{N}/N$, where $N=40$.  For the semi-analytic
model, we also consider a systematic uncertainty, that we summed in
quadrature to the Poissonian one.  Indeed, model predictions depend
mainly on two quantities: the effective cooling time of gas inside the
galactic dark matter potential wells, and the effectiveness of
interactions triggering powerful starbursts at high redshifts.  In the
model, both of the above processes can be slightly tuned to improve
the agreement with observations, provided that they remain consistent
with the observed color distribution of local galaxies. This degree of
freedom reflects into a factor $\sim 2.5$ systematic uncertainty
associated with the prediction of the abundance of passive galaxies at
high redshift ($z\geq 3$).

The model of \cite{menci14} predicts an overabundance of low-mass
passive galaxies and underestimates the massive tail of the SMF over
the entire redshift range probed by this work. This is a well known
problem of galaxy formation models
\citep{weinmann12,torrey14,white15,somerville15}: in a CDM scenario,
dwarf galaxies form first, hence are characterized by old stellar
populations, and feedback effects effectively suppress their star
formation. A partial solution to this problem has been implemented by
\cite{henriques17} through a revised feedback model tuned to fit the
observed passive fraction at $0<z<3$ (see also the comparison with the
SMF for passive galaxies shown by \citealt{girelli19}).

The overestimate of the low-mass tail is not observed in
ILLUSTRIS-TNG, where it has been suppressed by the introduction of a
minimum stellar wind velocity (the so-called wind velocity floor,
\citealt{pillepich18b,pillepich18a}).  The ILLUSTRIS-TNG300 simulation
performs in an excellent manner at $3<z<3.5$ and
$\log M/M_\odot\gtrsim 10.7$ in a still satisfactory way at higher
redshift and high ($\log M/M_\odot\gtrsim 11-11.5$) masses.  However,
it falls short of passive galaxies at low stellar masses compared to
the observations. \cite{donnari20} reported a very low fraction of
$M<10^{10.5}M_\odot$ passive galaxies predicted by ILLUSTRIS-TNG300 at
$0<z<3$. They concluded that the ILLUSTRIS-TNG model captures quite
well the effect of AGN feedback in quenching massive galaxies
(10.5$<$$\log
M/M_\odot$$<$11), but needs some adjustment for the feedback
mechanisms which regulate star formation in lower mass galaxies at
high redshift (e.g. stellar feedback).  The inability of reproducing
high-$z$ passive galaxies (with the exception of the highest masses)
is to be attributed to the still inefficient feedback.
ILLUSTRIS-TNG100 very nicely reproduces the SMF of passive galaxies at
$z<3.5$ and $\log M/M_\odot\gtrsim 10.7$, and it underpredicts the
number of passive galaxies at lower masses.  In the intermediate
(high) redshift interval, however, it only predicts the correct
abundance of passive sources at intermediate (high) stellar masses.
At the high mass end and intermediate redshifts, this could be due to
the paucity of such systems, that can be under-represented in small
simulated volumes (indeed, likely thanks to the larger volume,
ILLUSTRIS-TNG300 is able to reproduce the massive tail of the SMF in
the same redshift interval).  With the exceptions of the mass and
redshift bins where it predicts zero objects, ILLUSTRIS-TNG100 tends
to overall predict a higher SMF than ILLUSTRIS-TNG300. This trend is
also observed on the total SMF \citep{pillepich18b}, and it is due to
the combination of the limited numerical resolution (responsible for
the lower number of particles exceeding the fixed density threshold
above which stars can form) and volume effects.

An absence of massive passive galaxies as shown by TNG100 at
intermediate redshifts is observed in the EAGLE and SIMBA simulations,
both characterized by a similarly small volume. SIMBA predictions
confirm the trend already outlined in M19, where it turned out to be
the model showing the most serious underestimate of the number density
of passive galaxies at all redshifts above $\sim1-1.5$. While its
predictions are similar to those of ILLUSTRIS-TNG300 at $z>3.5$, it
underestimates the SMF also at $z<3.5$.

The most interesting comparison is however with EAGLE: it very nicely
reproduces the faint-end of the SMF, but it predicts no quiescent
galaxies at $M\geq 10^{11} M_\odot$. This was somewhat unexpected, as
in M19 we showed that EAGLE was the only hydrodynamic model, among
those considered, able to reproduce the number density of
$M>5 \times 10^9M_\odot$ passive galaxies at $z>3.5$ (an equally lack
of massive passive sources is obtained, as expected, when using the
more conservative cut sSFR$<$$10^{-11}$yr$^{-1}$,
as in M19). It implies that the number density alone is not a powerful
observable, as it does not encode information on the mass distribution
of the sources. This results suggests that EAGLE might reproduce the
abundance of high-$z$
passive galaxies thanks to an overprediction of lower mass
sources. Our data, however, do not allow to reliably measure the SMF
below $10^{10}M_\odot$ to test this hypothesis.
 
The lower panels of Fig.~\ref{fig:mfmodel} show the passive fraction
as a function of stellar mass. Some of the model analyzed match the
observations at low masses, others show a good match at high masses,
but none of them is able to reproduce the observed passive fraction
over the entire mass range.  The two ILLUSTRIS-TNG simulations are
consistent with the data in the lowest redshift interval, but are
below the observations at the low mass end. Similarly, they
underpredict the passive fraction at higher redshift, with the
exception of ILLUSTRIS-TNG100 that is consistent with the observations
at $3.5<z<4$
and $\log
M/M_\odot \sim
10.7$.  Conversely, over the entire redshift range probed, the
predictions of EAGLE and of the model of \cite{menci14} agree with the
observations at low masses, but no, or very few, passive galaxies are
predicted at high masses.  SIMBA predicts a very low fraction of
passive galaxies roughly over the entire mass range.

\subsubsection{The predicted SMD}

\begin{figure}
\centering
  \includegraphics[width=0.7\columnwidth,angle=270]{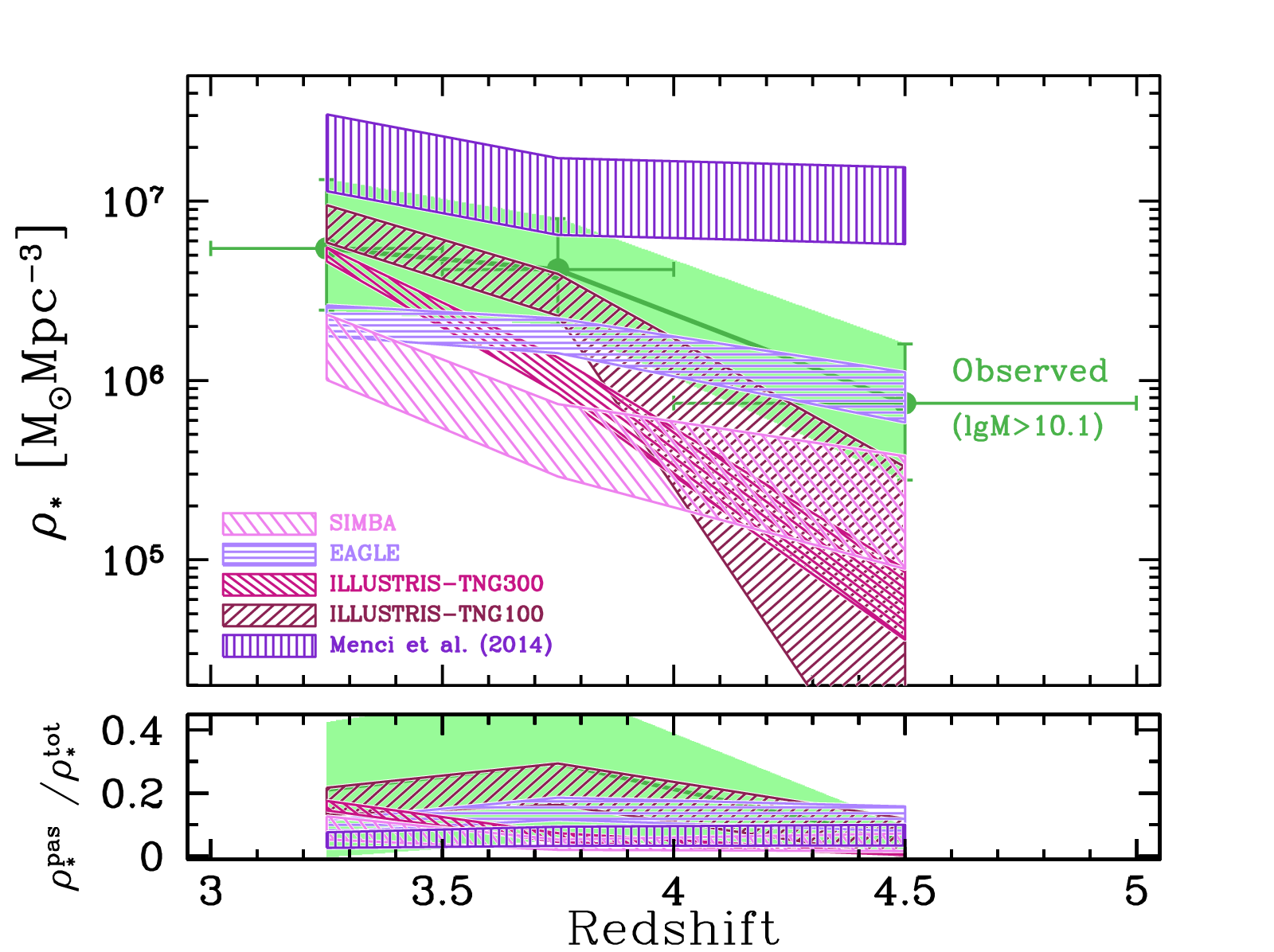}
  \caption{{\it Upper panel}: Evolution in the Stellar Mass Density of
    passive galaxies above $10^{10.1} M_\odot$ (green symbols and
    solid shaded area) compared to the theoretical predictions of the
    semi-analytic model of \cite{menci14} and the hydrodynamic
    simulations ILLUSTRIS-TNG100, ILLUSTRIS-TNG300, EAGLE and SIMBA
    (see legend), scaled to a Salpeter IMF.  {\it Lower panel}: Ratio
    between the mass density of passive galaxies and of the overall
    galaxy population above $10^{10.1} M_\odot$. The green solid line
    and solid region show the observed passive fraction, while the
    open shaded regions indicate model predictions, according to the
    legend.}
              \label{fig:mdmodel}
\end{figure}

Figure~\ref{fig:mdmodel} shows the comparison between the observed and
predicted SMD. For both models and data the SMD has been computed
above $10^{10.1}M_\odot$. The semi-analytic model of \cite{menci14}
overpredicts the observed SMD due to its steep shape at low masses,
dominating the integral.  Consistently with the comparison of the SMF,
the two ILLUSTRIS simulations reproduce the SMD of passive galaxies at
$z\sim3$, but underpredict it at higher redshifts. While the passive
SMD according to SIMBA is always below the observations, EAGLE is
marginally consistent at $z\sim3$ and nicely agrees with the data at
$z>4$. Indeed, despite its inability to reproduce the most massive
passive galaxies, the SMD is dominated by the low mass ones, and EAGLE
turns out to be the only model among those considered able to nicely
predict the faint-end of the passive SMF at the highest redshifts.
The lower panel of the same figure shows the ratio between the passive
and total SMD. ILLUSTRIS-TNG100 very nicely reproduces the observed
trend, and EAGLE is consistent within the uncertainties.
ILLUSTRIS-TNG300 and SIMBA predict a fraction of mass in passive
galaxies consistent with the observations at $z<3.5$.  Given the large
observational uncertainty, the model of \cite{menci14} as well as
ILLUSTRIS-TNG300 and SIMBA at $z>3.5$ are consistent with the data,
although they systematically predict a negligible fraction of mass
density accounted for by passive galaxies.

\subsubsection{Conclusions}

Overall, we observe a significant variation between the different
models, despite their predictions roughly agree for the local SMF. In
particular, the fact that none of the model analyzed is able to
reproduce the observed passive galaxies over the entire mass and
redshift range suggests that these passive derive from a subtle
combination of several physical processes.  Analyses like ours will
allow the validation of model recipes in a new and seemingly powerful
way. In any case, the inability of the models to correctly reproduce
the observed high-$z$ passive galaxies denotes a still incomplete
understanding of the physical mechanisms responsible for the formation
of these galaxies, their rapid assembly and abrupt suppression of
their star formation, and for galaxy evolution in general.

\section{Predictions for future surveys} \label{sec:predictions}

We extrapolated our SMFs to predict the intrinsic number of high-$z$
passive galaxies expected in surveys carried our with future
facilities, depending on their depth ($H$ band limiting magnitude) and
their area. Basically, from the typical observed mass-to-light ratio
of our candidates as a function of their observed magnitude, we
converted the limiting flux into a limiting stellar mass. We then
integrated the SMF at the different redshifts above this limiting mass
and rescaled to the relevant survey area.

We note that the number of galaxies predicted by extrapolating the SMF
is by definition larger than the number of candidates one will be able
to select due to incompleteness in the actual surveys.  As we discuss
below, the availability of ancillary photometry, especially longward
of the $H$ band, and the choice of the specific selection technique
will affect the actual number of candidates.

\begin{figure*}[!t]
    \centering
  \includegraphics[width=0.9\columnwidth,angle=270]{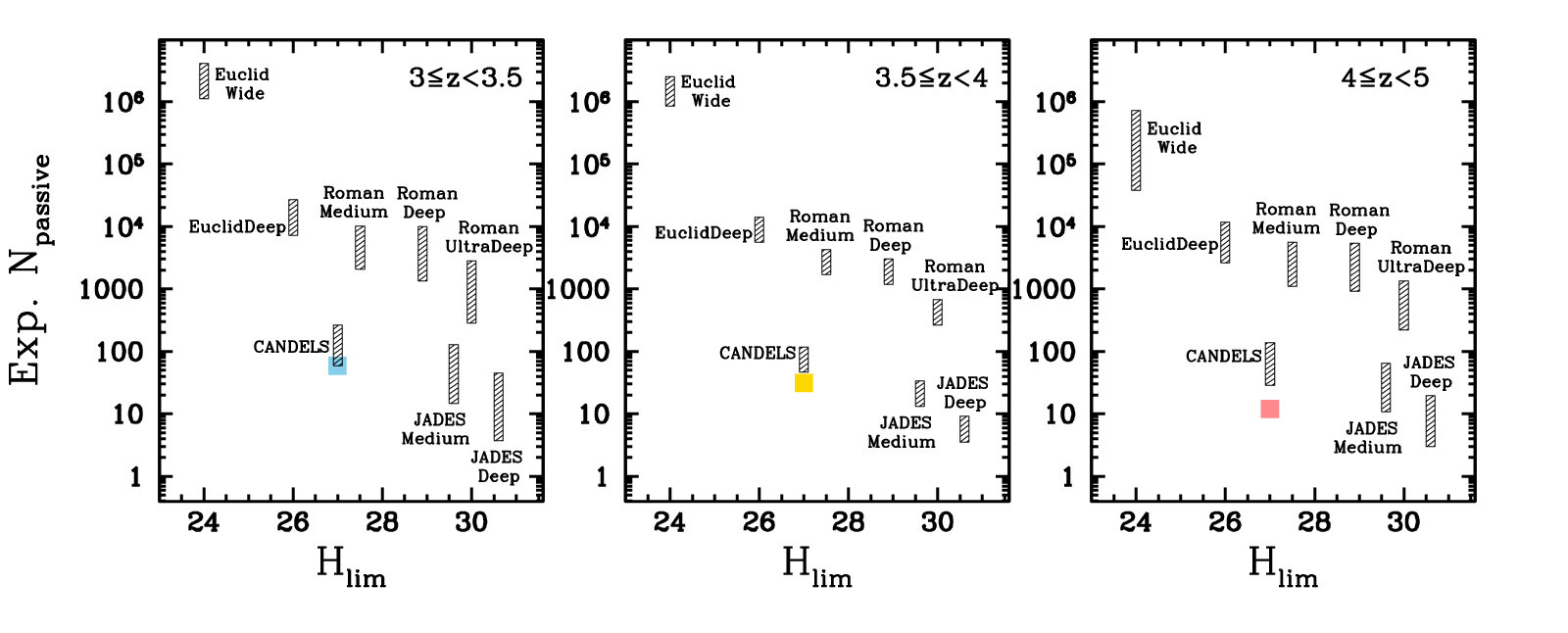}
  \caption[]{Predicted number of passive galaxies at different
    redshifts expected at different flux limits in the areas surveyed
    by future facilities (see Table~\ref{tab:predictions}).  The
    plotted range accounts for the uncertainty in the Schechter fits
    to the SMF (see text). The coloured boxes are the number of
    candidates actually observed in CANDELS (this work). We note that
    the predictions are by definition larger than the number of
    observed candidates due to incompleteness in the actual surveys.}
              \label{fig:predictions}
 \end{figure*}

\begin{table*}
\centering
\begin{tabular} {ccccccc}
\hline \hline 
\noalign{\smallskip} 
 Survey & Area [deg$^2$] & $H_{lim} (5\sigma)$& $\log M_{lim}/M_\odot$ & N$_{z=3-3.5}$ & N$_{z=3.5-4}$ & N$_{z=4-5}$\\
\noalign{\smallskip} 
\hline 
\noalign{\smallskip} 
JADES Deep & 0.013 &   30.6  &9.1  &  [5 -- 45]                           & [5 -- 10]                          & [5 -- 20] \\         
JADES Medium & 0.053 &  29.6 &9.4& [15 -- 130]                    & [15 -- 35]                       & [10 -- 65] \\        
Roman UltraDeep & 1 &  30    &9.3  & [30 -- 280]$\times 10^1$ &[25 -- 70]$\times 10^1$ & [20 -- 135]$\times 10^1$ \\         
Roman Deep & 5 &  28.9        &9.5  & [15 -- 100]$\times 10^2$  &[10 -- 30]$\times 10^2$&[10 -- 55]$\times 10^2$ \\         
Roman Medium & 9 &  27.5   &9.9  & [20 -- 100]$\times 10^2$  &[20 -- 45]$\times 10^2$ & [10 -- 55]$\times 10^2$\\         
Euclid Deep & 40 &  26        &10.3   & [70 -- 270]$\times 10^2$ &[55 -- 140]$\times 10^2$& [25 -- 120]$\times 10^2$\\         
Euclid Wide & 15000 &  24    &10.8  & [10 -- 40]$\times 10^5$ &[10 -- 25]$\times 10^5$ & [40 -- 725]$\times 10^3$ \\         
\noalign{\smallskip} \hline \noalign{\smallskip}
\end{tabular}
\caption{Predicted number of passive galaxies at different
  redshifts expected at different flux limits,  corresponding to
  different mass limits, in the areas surveyed
  by future facilities. The range in square brackets encompasses the
  1$\sigma$ uncertainty in the predicted number caused by
  the uncertainties 
  in the Schechter fits to the SMF (see text). 
}\label{tab:predictions}
\end{table*}

The calculation is based on two main assumptions. The first one is
that the observed mass-to-light ratio ($\log M - H$) vs $H$ band
magnitude relation, well fitted by a linear slope at magnitudes
brighter than 27, shows the same trend at fainter values. The second
assumption is that the SMF, that we measure above $10^{10}M_\odot$,
can be safely extrapolated to lower masses (we note that the faint-end
slope was only fitted to the data in our lowest redshift bin, and it
was kept fixed at higher redshifts). Both assumptions affect the
predictions at magnitude deeper than 27, which is the magnitude cut of
our selection. Indeed, $H$=27 roughly corresponds to $10^{10}M_\odot$,
where our SMFs are still safely based on the observations and do not
suffer from strong uncertainties associated with their
extrapolation. Predictions for surveys deeper than $H$=27 should
therefore been taken with caution.
 
Another important issue to be taken into account is that the selection
function for high-$z$ passive galaxies is not simply based on the
depth of the observations in the $H$ band, but it depends in a complex
way on the depth in other filters, in particular the $K$ and IRAC CH1
and CH2 bands, as described in Sect.\ref{sec:passivesel} and fully
discussed in M19. This implies that, although future telescopes are
expected to survey large sky areas down to very deep limiting $H$ band
fluxes, additional data will be needed to properly select the high-$z$
passive galaxies predicted to be found in the observed sky
regions. The selection efficiency will indeed crucially depend on the
availability of ancillary data.  Our predictions can therefore be
considered as upper limits to the expected number of candidates that
will be detected by the future telescopes under consideration.

Keeping all these caveats in mind, the results of our calculation are
shown in Fig.~\ref{fig:predictions} and reported in
Table~\ref{tab:predictions}.  Given the large uncertainties associated
with our SMFs, we only report the 68\% confidence level intervals on
the predicted number of passive galaxies.  Since the slope of the SMF
has been fixed at $z>3.5$, the associated confidence regions are
narrower than at lower redshift, and so are the uncertainties on our
predictions.

We first compared the expected number of passive sources in an area
equal to the CANDELS one down to $H$=27 to the number of candidates
actually observed. As expected, the number of selected passive
galaxies is lower than our predictions by a factor of $\sim 2-5$.  As
discussed above, this mismatch is explained by the incompleteness in
the observed sample. It is larger at higher redshift due to the higher
level of incompleteness, for a given mass limit.

Besides a CANDELS-like survey, we considered the JWST Advanced Deep
Extragalactic Survey (JADES), Euclid and the Nancy Grace Roman Space
Telescope. JADES\footnote{https://www.cosmos.esa.int/web/jwst-nirspec-gto/jades}
is a JWST GTO program, imaging with NIRCam 46 and 190 arcmin$^2$ down
to a limit of AB=30.6 and 29.6, respectively for the Deep and Medium
survey, in the GOODS-S and GOODS-N fields. The Euclid Deep and Wide
surveys will scan 40 and 15000 deg$^2$ down to a limiting depth of
$H$=24 and $H$=26 \citep{laureijs11}, respectively. For the Nancy
Grace Roman Space Telescope, we considered an UltraDeep survey
reaching $H$=30 over 1 deg$^2$ \citep{koekemoer19} and approximate
depths of $\sim$28.9 and $\sim$27.5 over $\sim$5 and $\sim$9 deg$^2$
for a Deep and a Medium survey, respectively \citep{hounsell18}.

Based on our predictions, given the scarcity of passive galaxies in
the young Universe, the surveyed sky area turns out to be the key
feature to perform statistical studies of this class of sources, more
than the observing depth. Thanks to its much larger area, the Euclid
Deep survey, despite expected to be 1.5 and 3 magnitude shallower,
will provide more candidates than the Medium and Deep surveys,
respectively, carried out with the Nancy Grace Roman Space Telescope.
The Euclid Wide survey will further improve the statistics by a factor
of $\sim$100, and by four orders of magnitude compared to what is now
possible. We remind, however, that we are simply predicting the number
of candidates expected in a given sky area down to a given flux limit,
and not our capability in selecting them. As discussed above,
ancillary, deep enough data are needed at longer wavelength to
complement optical/near-IR observations.  Spitzer data from the SIMPLE
survey \citep{damen11} over the Extended CDF-South ($\sim$ 0.5
deg$^2$) and especially from the Euclid/WFIRST Spitzer Legacy Survey
(covering 20 deg$^2$, proposals IDs \#13058 and \#13153, PI P. Capak)
will be of great help, but will not cover the entire sky areas that
will be observed by Euclid and Roman.  This will allow us to only
detect a fraction of the predicted number of high-$z$ passive
candidates.

Despite its small field of view, hence the relatively small areas
surveyed, in the next future JWST will be the only instrument that
will allow a self-standing selection of high-$z$ passive galaxies,
thanks to its red filters ($> 2\mu$m) that will complement the $H$
band observations to a similar depth. JWST will allow a much cleaner
selection of the candidates \citep[see][]{merlin18}, an easier and
faster spectroscopical confirmation, and it will extend the selection
to even higher redshift thanks to its redder filters compared to
CANDELS.

\section{Summary} \label{sec:summary}

In this paper, we present a follow-up analysis of the passive galaxy
candidates selected at $z>3$ in the CANDELS fields by M19 through an
accurate and {\it ad-hoc} developed SED fitting technique. After
confirming their passive nature by means of their sub-mm emission,
probing the lack of on-going star formation, we study the Stellar Mass
Function (SMF) of the passive population at $3<z<5$. We summarize our
results in the following.

We started by searching the ALMA archive for observations of our
candidates. We found data for 77\% of the sources located in the
fields accessible by ALMA. Following the method presented in our
previous work (S19), based on the comparison between ALMA predictions
and the outcome of the SED fitting analysis, we could confirm the
accuracy of the passive classification for 61\% of the candidates on
an individual basis at a 3$\sigma$ confidence level.  Since the
available data is not deep enough to confirm or reject the rest of the
candidates, we validated the population as being on average passive in
a statistical sense, through stacking and by comparison with the
location of the Main Sequence of star-forming galaxies at the same
redshifts. 

Once again (see S19), we confirmed the reliability and robustness of
the photometric selection technique developed in
\cite{merlin18}. Assuming that it performs equally well on the sources
for which no sub-mm information is available, we computed the SMF over
the entire sample of 101 $3<z<5$ passive candidates using a stepwise
approach.  The latter has the advantage of taking into account
photometric errors, mass completeness issues and the Eddington bias,
without any a-posteriori correction.

We analysed the evolution in the SMF from $z=5$ to $z=3$ and compared
it with the SMF of the total population. Despite the large associated
uncertainties, we observe a strong evolution in the SMF around
$z\sim 4$, indicating that we are witnessing the emergence of the
passive population at this epoch, i.e. we are looking at the epoch at
which these galaxies become passive. Consistently, we observe an
increase in the abundance of passive galaxies compared to the overall
population at the same redshift. While quiescent galaxies are always
below $\lesssim 10\%$ at low ($M\lesssim 10^{10.5}M_\odot$) stellar
masses, they can make a large fraction of the entire population (up to
or more than 50\%) at $M>10^{11}M_\odot$ and $z<4$. On the other hand,
their abundance remains low ($\lesssim  10\%$) at $z>4$. Integrating
over $M>10^{10} M_\odot$, passive galaxies account for $\sim 20-25\%$ of
the total stellar mass density at $z=3-4$, and only $\sim 5\%$ at
earlier epochs. Current uncertainties on these numbers are however
very large and prevent more accurate conclusions.
 
We compared our results with the literature, and found a factor of 4
to 10 more passive sources than most previous observations, with the
exception of a couple of works consistent with our results within the
uncertainties, at least in some of the redshift bins. This is due to a
combination of {\it a)} the better quality of the CANDELS
observations, {\it b)} the very accurate selection technique developed
by \cite{merlin18}, that include recently quenched galaxies often
missed by colour-colour selections, and {\it c)} the method adopted to
calculate the SMF, that intrinsically accounts for mass and selection
incompleteness, as well as for the Eddington bias.  We also compared
our results with theoretical predictions. We found an overall
agreement with at least some of the models considered, but also clear
mismatches, denoting a still incomplete understanding of the physical
processes responsible for the formation of these galaxies and for
galaxy evolution in general.

The analysis of passive galaxies in the high redshift Universe will be
greatly improved by future facilities. JWST observations will improve
their selection, and thanks to its spectroscopical capabilities will
allow a faster confirmation as well as a deeper investigation of their
physical processes. On the other hand, wide field surveys with Euclid
and the Nancy Grace Roman Space Telescope, complemented with ancillary
optical and IR data, will be crucial to improve the statistical
analysis of these rare sources.  Depth, redder filters and area will
all combine to reduce the uncertainties currently affecting this kind
of studies.

\begin{acknowledgements}
  We thank the anonymous referee for the thorough and careful review
  of the manuscript.  This paper makes use of the following ALMA data:
  2012.1.00869.S, 2012.1.00173.S, 2013.1.00718.S, 2013.1.01292.S,
  2013.1.00118.S, 2015.1.00242.S, 2015.1.00543.S, 2015.1.01074.S,
  2015.1.00098.S, 2015.1.00870.S, 2015.1.00379.S, 2015.1.01495.S,
  2016.1.01079.S. The IllustrisTNG simulations were undertaken with
  compute time awarded by the Gauss Centre for Supercomputing (GCS)
  under GCS Large-Scale Projects GCS-ILLU and GCS-DWAR on the GCS
  share of the supercomputer Hazel Hen at the High Performance
  Computing Center Stuttgart (HLRS), as well as on the machines of the
  Max Planck Computing and Data Facility (MPCDF) in Garching,
  Germany. We acknowledge the Virgo Consortium for making their
  simulation data available. The EAGLE simulations were performed
  using the DiRAC-2 facility at Durham, managed by the ICC, and the
  PRACE facility Curie based in France at TGCC, CEA,
  Bruy\`eres-le-Ch\^atel. We also thank Romeel Dav\'e for help in
  using
the SIMBA simulation data.
\end{acknowledgements}

\nocite{tomczak14}

\bibliographystyle{aa}

 \begin{appendix}

\section{List of passive galaxy candidates} \label{sec:list}

We list in Tables~\ref{tab:gs} to \ref{tab:egs} the passive candidates
selected by \cite{merlin19} and used in this work (we do not include
one source at $z=6.7$ as outside the redshift range probed) to compute
the passive SMF.  We report their redshift, $H$ band magnitude and
stellar masses. Tables~\ref{tab:gs}, \ref{tab:uds} and
\ref{tab:cosmos} show the candidates in GOODS-S, UDS and COSMOS,
respectively, and also report the sub-mm fluxes and inferred (limits
on the) SFR for the sources having ALMA archival observations in Band
6 and 7. Tables~\ref{tab:gn} and \ref{tab:egs} are for the GOODS-N and
EGS fields, unobservable with ALMA.  We refer the reader to Table~B.1
of \cite{merlin19} for candidate positions, best-fit ages and time
since quiescence.

\begin{table*}
\centering
\begin{tabular} {ccccccccc}
\hline 
\hline 
\noalign{\smallskip} 
 ID & $z$ & $H$ & Stellar mass  & ALMA band & ALMA flux & SFR$_{M10}$ & SFR$_{Ma21}$ &  Confirmed\\
& & [mag] & $\log$ ($M$/$M_\odot$) & & [mJy/beam] & [$M_\odot$/$yr$] & [$M_\odot$/$yr$] &  (Y/N)\\
\noalign{\smallskip} 
\hline 
\noalign{\smallskip} 
GOODSS-2608 & 3.72 & 26.03 &  9.58$_{-0.06}^{+0.11}$ & 7 &  0.01$ \pm $0.13 & $<$20 & $<$5 & N\\
\noalign{\smallskip} 
GOODSS-2717 & 3.03 & 23.98 & 11.12$_{-0.08}^{+0.19}$ & -- & -- & -- & --& --  \\
\noalign{\smallskip} 
GOODSS-2782 & 3.58 & 24.94 & 10.86$_{-0.05}^{+0.04}$ & 7 &  0.05$ \pm $0.11 & $<$23 & $<$5 & Y\\
\noalign{\smallskip} 
GOODSS-3718 & 3.85 & 24.91 & 10.33$_{-0.01}^{+0.06}$ & -- & -- & -- & -- & -- \\
\noalign{\smallskip} 
GOODSS-3897 & 3.12 & 24.64 & 10.28$_{-0.06}^{+0.01}$ & 7 & -0.05$ \pm $0.50 & $<$74 & $<$16  & Y\\
\noalign{\smallskip} 
GOODSS-3912 & 3.90 & 26.86 & 10.55$_{-0.13}^{+0.09}$ & 6 & -0.07$ \pm $0.17 & $<$43 & $<$8  & N\\
\noalign{\smallskip} 
GOODSS-3973 & 3.63 & 25.42 & 11.23$_{-0.10}^{+0.08}$ & 7 &  0.65$ \pm $0.18 & 95$ \pm $26 & 23$ \pm $6  & Y\\
\noalign{\smallskip} 
GOODSS-4202 & 3.31 & 26.06 &  9.66$_{-0.15}^{+0.04}$ & -- & -- & -- & -- & -- \\
\noalign{\smallskip} 
GOODSS-4503 & 3.59 & 24.07 & 11.14$_{-0.07}^{+0.03}$ & 7 & -0.03$ \pm $0.29 & $<$42 & $<$10  & Y\\
\noalign{\smallskip} 
GOODSS-4587 & 3.75 & 25.84 &  9.74$_{-0.07}^{+0.12}$ & 6 & -0.08$ \pm $0.17 & $<$43 & $<$8  & N\\
\noalign{\smallskip} 
GOODSS-4949 & 4.82 & 25.54 & 10.00$_{-0.01}^{+0.03}$ & 6 & -0.07$ \pm $0.23 & $<$55 & $<$13  & N\\
\noalign{\smallskip} 
GOODSS-5934 & 4.86 & 25.81 &  9.83$_{-0.02}^{+0.01}$ & 6 &  0.13$ \pm $0.17 & $<$72 & $<$17  & N\\
\noalign{\smallskip} 
GOODSS-6407 & 4.81 & 25.32 & 10.23$_{-0.07}^{+0.08}$ & 6 & -0.11$ \pm $0.17 & $<$42 & $<$10  & N\\
\noalign{\smallskip} 
GOODSS-7526 & 3.32 & 25.98 & 10.54$_{-0.25}^{+0.10}$ & 7 &  0.11$ \pm $0.12 & $<$34 & $<$7  & Y\\
\noalign{\smallskip} 
GOODSS-7688 & 3.40 & 25.83 & 10.37$_{-0.24}^{+0.08}$ & 6 &  0.23$ \pm $0.22 & 58$ \pm $56 & 11$ \pm $10  & Y\\
\noalign{\smallskip} 
GOODSS-8242 & 3.24 & 25.31 &  9.84$_{-0.01}^{+0.01}$ & 6 &  0.17$ \pm $0.23 & $<$102 & $<$19 & N\\
\noalign{\smallskip} 
GOODSS-8785 & 3.85 & 26.50 & 10.55$_{-0.23}^{+0.13}$ & 6 & -0.37$ \pm $0.22 & $<$56 & $<$11 & N\\
\noalign{\smallskip} 
GOODSS-9209 & 4.49 & 24.75 & 10.96$_{-0.01}^{+0.01}$ & 6 & -0.05$ \pm $0.16 & $<$41 & $<$9  & Y\\
\noalign{\smallskip} 
GOODSS-10578 & 3.06 & 22.63 & 11.52$_{-0.19}^{+0.01}$ & 6 &  0.13$ \pm $0.09 & 35$ \pm $24 & 6$ \pm $4  & Y\\
\noalign{\smallskip} 
GOODSS-12178 & 3.29 & 25.15 & 10.68$_{-0.12}^{+0.08}$ & 6 &  0.13$ \pm $0.17 & $<$77 & $<$14  & Y\\
\noalign{\smallskip} 
GOODSS-13394 & 3.29 & 24.90 & 10.06$_{-0.04}^{+0.01}$ & 6 &  0.06$ \pm $0.22 & $<$72 & $<$13   & Y\\
\noalign{\smallskip} 
GOODSS-15457 & 3.50 & 25.56 &  9.67$_{-0.06}^{+0.06}$ & 6 &  0.02$ \pm $0.03 & $<$16 & $<$3  & N\\
\noalign{\smallskip} 
GOODSS-16506 & 3.38 & 25.44 &  9.72$_{-0.02}^{+0.13}$ & 6 &  0.01$ \pm $0.02 & $<$8 & $<$1  & N\\
\noalign{\smallskip} 
 GOODSS-16526 & 3.15 & 24.62 & 10.13$_{-0.01}^{+0.01}$ & 7 &  0.00$ \pm $0.34 & $<$50 & $<$11 & Y\\
\noalign{\smallskip} 
GOODSS-17749 & 3.70 & 25.25 & 10.98$_{-0.06}^{+0.10}$ & 6 &  0.11$ \pm $0.08 & 29$ \pm $20 & 5$ \pm $4  & Y\\
\noalign{\smallskip} 
GOODSS-18180 & 3.65 & 25.13 & 10.91$_{-0.04}^{+0.09}$ & 6 &  0.08$ \pm $0.08 & 21$ \pm $20 & 4$ \pm $3   & Y\\
\noalign{\smallskip} 
GOODSS-19301 & 3.59 & 26.38 & 10.13$_{-0.18}^{+0.06}$ & 6 &  0.01$ \pm $0.08 & $<$25 & $<$5   & Y\\
\noalign{\smallskip} 
GOODSS-19446 & 3.27 & 24.48 & 10.25$_{-0.01}^{+0.01}$ & 6 &  0.04$ \pm $0.09 & $<$33 & $<$6   & Y\\
\noalign{\smallskip} 
GOODSS-19505 & 3.59 & 24.36 & 10.72$_{-0.01}^{+0.01}$ & 6 &  0.00$ \pm $0.04 & $<$14 & $<$2  & Y\\
\noalign{\smallskip} 
GOODSS-19883 & 3.57 & 24.50 & 11.20$_{-0.06}^{+0.03}$ & 7 &  0.30$ \pm $0.25 & 44$ \pm $36 & 10$ \pm $9  & Y\\
\noalign{\smallskip} 
GOODSS-22085 & 3.47 & 25.13 & 10.68$_{-0.08}^{+0.05}$ & -- & -- & -- & -- & -- \\
\noalign{\smallskip} 
GOODSS-22610 & 3.33 & 24.81 & 10.05$_{-0.03}^{+0.06}$ & 7 &  0.08$ \pm $0.24 & $<$46 & $<$10  & N\\
\noalign{\smallskip} 
GOODSS-23626 & 4.75 & 25.57 & 10.89$_{-0.06}^{+0.06}$ & 7 &  0.17$ \pm $0.30 & $<$76 & $<$23  & N\\
\noalign{\smallskip} 
\hline 
\end{tabular}
\caption{Passive galaxy candidates at $3<z<5$ in the GOODS-S field.
}  
\label{tab:gs}
\end{table*}

\begin{table*}
\centering
\begin{tabular} {ccccccccc}
\hline 
\hline 
\noalign{\smallskip} 
 ID & $z$ & $H$ & Stellar mass  & ALMA band & ALMA flux & SFR$_{M10}$ & SFR$_{Ma21}$ &Confirmed\\
& & [mag] & $\log$ ($M$/$M_\odot$) & & [mJy/beam] & [$M_\odot$/$yr$] & [$M_\odot$/$yr$]  & (Y/N)\\
\noalign{\smallskip} 
\hline 
\noalign{\smallskip} 
UDS-1244 & 3.79 & 24.80 & 11.14$_{-0.12}^{+0.01}$ & 7 & -0.17$ \pm $0.36 & $<$52 & $<$13 &N\\
\noalign{\smallskip} 
UDS-2571 & 3.70 & 25.14 & 10.84$_{-0.13}^{+0.07}$ & 7 & -0.05$ \pm $0.55 & $<$81 & $<$20 & N \\
\noalign{\smallskip} 
UDS-4332 & 3.18 & 24.78 & 11.25$_{-0.07}^{+0.17}$ & 7 & -0.07$ \pm $0.24 & $<$36 & $<$7  & Y\\
\noalign{\smallskip} 
UDS-7520 & 3.16 & 24.28 & 11.44$_{-0.09}^{+0.01}$ & -- & -- & -- & -- & -- \\
\noalign{\smallskip} 
UDS-7779 & 3.14 & 24.42 & 11.01$_{-0.12}^{+0.17}$ & -- & -- & -- & -- & -- \\
\noalign{\smallskip} 
UDS-8682 & 3.46 & 25.11 & 10.87$_{-0.13}^{+0.09}$ & -- & -- & -- & -- & -- \\
\noalign{\smallskip} 
UDS-8689 & 3.22 & 23.51 & 11.24$_{-0.01}^{+0.02}$ & -- & -- & -- & -- & -- \\
\noalign{\smallskip} 
UDS-10086 & 3.09 & 24.63 & 10.50$_{-0.05}^{+0.09}$ & -- & -- & -- & --& --  \\
\noalign{\smallskip} 
UDS-10430 & 4.13 & 25.54 & 11.02$_{-0.14}^{+0.11}$ & 7 &  0.03$ \pm $0.26 & $<$42 & $<$11 & N\\
\noalign{\smallskip} 
UDS-11532 & 4.21 & 24.93 & 10.25$_{-0.08}^{+0.09}$ & -- & -- & -- & -- & -- \\
\noalign{\smallskip} 
UDS-12640 & 3.61 & 25.05 & 10.66$_{-0.07}^{+0.17}$ & 7 &  0.00$ \pm $0.34 & $<$50 & $<$12 & Y\\
\noalign{\smallskip} 
UDS-20843 & 3.73 & 24.16 & 10.98$_{-0.01}^{+0.06}$ & 7 & -0.04$ \pm $0.22 & $<$31 & $<$8  & Y\\
\noalign{\smallskip} 
UDS-23628 & 4.25 & 24.67 & 10.92$_{-0.01}^{+0.14}$ & 7 &  0.05$ \pm $0.35 & $<$60 & $<$17  & Y\\
\noalign{\smallskip} 
UDS-25688 & 3.08 & 23.10 & 11.42$_{-0.02}^{+0.02}$ & 7 &  0.22$ \pm $0.24 & $<$68 & $<$14  & Y\\
\noalign{\smallskip} 
UDS-25893 & 4.49 & 26.32 & 11.37$_{-0.08}^{+0.10}$ & 7 &  0.01$ \pm $0.34 & $<$54 & $<$16 &N\\
\noalign{\smallskip} 
UDS-32406 & 3.28 & 26.10 & 10.27$_{-0.06}^{+0.09}$ & -- & -- & -- & -- & -- \\
\noalign{\smallskip} 
\hline 
\end{tabular}
\caption{Passive galaxy candidates at $3<z<5$ in the UDS field.
}  
\label{tab:uds}
\end{table*}

\begin{table*}
\centering
\begin{tabular} {ccccccccc}
\hline 
\hline 
\noalign{\smallskip} 
 ID & $z$ & $H$ & Stellar mass  & ALMA band & ALMA flux & SFR$_{M10}$  & SFR$_{Ma21}$ & Confirmed\\
& & [mag] & $\log$ ($M$/$M_\odot$) & & [mJy/beam] & [$M_\odot$/$yr$] & [$M_\odot$/$yr$] & (Y/N)\\
\noalign{\smallskip} 
\hline 
\noalign{\smallskip} 
COSMOS-2075 & 3.35 & 24.43 & 10.67$_{-0.14}^{+0.07}$ & 6 & -0.04$ \pm $0.06 & $<$20 & $<$3 &Y \\
\noalign{\smallskip} 
COSMOS-16676 & 3.72 & 24.60 & 11.51$_{-0.14}^{+0.16}$ & 7 &  0.50$ \pm $0.20 & 72$ \pm $29 & 18$ \pm $7 &Y\\
\noalign{\smallskip} 
COSMOS-18286 & 3.04 & 23.89 & 11.37$_{-0.12}^{+0.01}$ & -- & -- & -- & -- & -- \\
\noalign{\smallskip} 
COSMOS-19502 & 3.87 & 24.46 & 11.07$_{-0.09}^{+0.12}$ & 7 &  0.04$ \pm $0.18 & $<$32 & $<$8 &Y\\
\noalign{\smallskip} 
\hline 
\end{tabular}
\caption{Passive galaxy candidates at $3<z<5$ in the COSMOS field.
}  
\label{tab:cosmos}
\end{table*}

\begin{table*}
\centering
\begin{tabular} {cccc}
\hline 
\hline 
\noalign{\smallskip} 
 ID & $z$ & $H$ & Stellar mass  \\
& & [mag] & $\log$ ($M$/$M_\odot$) \\
\noalign{\smallskip} 
\hline 
\noalign{\smallskip} 
GOODSN-13 & 3.01 & 23.87 & 10.04$_{-0.01}^{+0.03}$ \\
\noalign{\smallskip} 
GOODSN-357 & 3.09 & 24.15 &  9.98$_{-0.06}^{+0.10}$   \\
\noalign{\smallskip} 
GOODSN-1570 & 3.23 & 23.48 & 10.12$_{-0.01}^{+0.08}$   \\
\noalign{\smallskip} 
GOODSN-2901 & 3.70 & 25.30 & 10.07$_{-0.16}^{+0.23}$   \\
\noalign{\smallskip} 
GOODSN-4004 & 3.81 & 25.56 & 10.56$_{-0.06}^{+0.08}$   \\
\noalign{\smallskip} 
GOODSN-4691 & 3.18 & 24.75 & 11.04$_{-0.15}^{+0.18}$   \\
\noalign{\smallskip} 
GOODSN-5059 & 3.69 & 25.50 & 11.10$_{-0.19}^{+0.12}$   \\
\noalign{\smallskip} 
GOODSN-5744 & 3.46 & 25.81 & 10.70$_{-0.17}^{+0.16}$   \\
\noalign{\smallskip} 
GOODSN-6430 & 3.21 & 24.86 &  9.94$_{-0.07}^{+0.01}$   \\
\noalign{\smallskip} 
GOODSN-6620 & 3.70 & 23.87 & 10.44$_{-0.01}^{+0.01}$   \\
\noalign{\smallskip} 
GOODSN-7385 & 3.18 & 25.09 &  9.90$_{-0.15}^{+0.06}$   \\
\noalign{\smallskip} 
GOODSN-9626 & 3.18 & 24.71 &  9.89$_{-0.05}^{+0.07}$   \\
\noalign{\smallskip} 
GOODSN-10956 & 3.09 & 24.90 &  9.64$_{-0.01}^{+0.16}$   \\
\noalign{\smallskip} 
GOODSN-11579 & 3.17 & 25.63 & 10.73$_{-0.18}^{+0.16}$   \\
\noalign{\smallskip} 
GOODSN-12446 & 3.05 & 25.40 &  9.29$_{-0.01}^{+0.01}$   \\
\noalign{\smallskip} 
GOODSN-13007 & 3.04 & 24.18 & 10.57$_{-0.21}^{+0.06}$   \\
\noalign{\smallskip} 
GOODSN-13403 & 3.79 & 26.14 &  9.81$_{-0.16}^{+0.09}$   \\
\noalign{\smallskip} 
GOODSN-13435 & 3.65 & 25.65 & 10.79$_{-0.17}^{+0.13}$   \\
\noalign{\smallskip} 
GOODSN-13800 & 3.33 & 23.72 & 10.87$_{-0.05}^{+0.06}$   \\
\noalign{\smallskip} 
GOODSN-14482 & 3.49 & 25.30 & 10.05$_{-0.04}^{+0.10}$   \\
\noalign{\smallskip} 
GOODSN-15054 & 3.06 & 22.90 & 11.32$_{-0.01}^{+0.04}$   \\
\noalign{\smallskip} 
GOODSN-16817 & 3.69 & 24.78 &  9.81$_{-0.01}^{+0.10}$   \\
\noalign{\smallskip} 
GOODSN-18860 & 4.53 & 24.95 & 10.11$_{-0.01}^{+0.12}$   \\
\noalign{\smallskip} 
GOODSN-19580 & 3.10 & 24.24 & 11.03$_{-0.04}^{+0.09}$   \\
\noalign{\smallskip} 
GOODSN-20589 & 3.34 & 24.85 &  9.88$_{-0.01}^{+0.11}$   \\
\noalign{\smallskip} 
GOODSN-21034 & 3.33 & 26.19 &  9.98$_{-0.10}^{+0.07}$   \\
\noalign{\smallskip} 
GOODSN-21961 & 3.36 & 25.52 & 10.44$_{-0.17}^{+0.13}$   \\
\noalign{\smallskip} 
GOODSN-22398 & 3.11 & 24.87 &  9.67$_{-0.11}^{+0.01}$   \\
\noalign{\smallskip} 
GOODSN-24092 & 3.30 & 23.99 & 11.09$_{-0.03}^{+0.06}$   \\
\noalign{\smallskip} 
GOODSN-24501 & 4.24 & 24.93 & 10.32$_{-0.03}^{+0.17}$   \\
\noalign{\smallskip} 
GOODSN-24572 & 3.34 & 23.96 & 10.22$_{-0.07}^{+0.05}$   \\
\noalign{\smallskip} 
GOODSN-25209 & 3.27 & 24.65 & 10.79$_{-0.12}^{+0.05}$   \\
\noalign{\smallskip} 
GOODSN-27251 & 3.12 & 24.59 & 10.07$_{-0.07}^{+0.01}$   \\
\noalign{\smallskip} 
GOODSN-28344 & 4.76 & 26.72 & 10.44$_{-0.17}^{+0.14}$   \\
\noalign{\smallskip} 
GOODSN-35028 & 3.64 & 25.17 & 11.24$_{-0.15}^{+0.08}$   \\
\noalign{\smallskip} 
\hline 
\end{tabular}
\caption{Passive galaxy candidates at $3<z<5$ in the GOODS-N field.
}  
\label{tab:gn}
\end{table*}

\begin{table*}
\centering
\begin{tabular} {cccc}
\hline 
\hline 
\noalign{\smallskip} 
 ID & $z$ & $H$ & Stellar mass  \\
& & [mag] & $\log$ ($M$/$M_\odot$) \\
\noalign{\smallskip} 
\hline 
\noalign{\smallskip} 
 EGS-2490 & 3.10 & 24.41 & 10.59$_{-0.15}^{+0.02}$   \\
 \noalign{\smallskip} 
 EGS-6539 & 3.44 & 23.84 & 11.14$_{-0.12}^{+0.01}$   \\
 \noalign{\smallskip} 
 EGS-14727 & 3.05 & 22.88 & 11.19$_{-0.04}^{+0.09}$   \\
 \noalign{\smallskip} 
 EGS-15868 & 3.61 & 23.61 & 11.02$_{-0.01}^{+0.11}$   \\
 \noalign{\smallskip} 
 EGS-21351 & 3.61 & 25.03 & 10.91$_{-0.06}^{+0.06}$   \\
 \noalign{\smallskip} 
 EGS-23036 & 3.57 & 25.48 & 10.57$_{-0.09}^{+0.04}$   \\
 \noalign{\smallskip} 
 EGS-24177 & 3.42 & 23.58 & 11.33$_{-0.06}^{+0.07}$   \\
 \noalign{\smallskip} 
 EGS-24356 & 3.43 & 24.30 & 11.04$_{-0.06}^{+0.05}$   \\
 \noalign{\smallskip} 
 EGS-25724 & 3.80 & 25.34 & 10.89$_{-0.07}^{+0.09}$   \\
 \noalign{\smallskip} 
 EGS-26762 & 3.28 & 24.92 & 10.29$_{-0.11}^{+0.09}$   \\
 \noalign{\smallskip} 
 EGS-27491 & 3.34 & 24.35 & 10.99$_{-0.13}^{+0.05}$   \\
 \noalign{\smallskip} 
 EGS-29547 & 3.14 & 24.30 & 10.94$_{-0.09}^{+0.06}$   \\
 \noalign{\smallskip} 
 EGS-30675 & 3.01 & 24.43 & 10.70$_{-0.15}^{+0.03}$   \\
\noalign{\smallskip} 
\hline 
\end{tabular}
\caption{Passive galaxy candidates at $3<z<5$ in the EGS field.
}  
\label{tab:egs}
\end{table*}

\end{appendix}

\end{document}